\documentclass[ALICE,manyauthors]{cernphprep}

\newcommand{\pp}{p\kern-0.05em p}
\newcommand{\ppbar}{\mathrm{p}\kern-0.05em \overline{\mathrm{p}}}

\newcommand{\GeV}{\ensuremath{\mathrm{GeV}\kern-0.05em}}
\newcommand{\GeVc}{\ensuremath{\mathrm{GeV}\kern-0.05em/\kern-0.02em c}}

\usepackage{diagbox}

\newcommand{\pTchjet}{\ensuremath{p_{\mathrm{T}}^{\mathrm{ch\; jet}}}}

\usepackage[comma,square,numbers,sort&compress]{natbib}
\usepackage{hyperref}
\usepackage{lineno}
\usepackage{amsmath}
\usepackage{booktabs}
\usepackage{bigints}  % larger integral signs
\usepackage{afterpage}  % big figures can have their own page
\usepackage{multirow}
\usepackage{xcolor}
\usepackage[Symbol]{upgreek}
\usepackage[T1]{fontenc}
\usepackage{orcidlink}
\usepackage{mathtools}
\usepackage{siunitx}
\usepackage{soul}
\DeclarePairedDelimiterXPP\BigOSI[2]%
  {\mathcal{O}}{(}{)}{}%
  {\SI{#1}{#2}}
\usepackage{xspace}
\usepackage{color}
\usepackage{xcolor}
\usepackage{rotating}
\definecolor{light-gray}{gray}{0.8}

\newcommand{\TeV}{\ensuremath{\rm TeV}\xspace}

\newcommand{\sqrtsNN}{\ensuremath{\sqrt{s_{\rm NN}}}\xspace}
\newcommand{\pt}{\ensuremath{{\it p}_{\rm T}}\xspace}
\newcommand{\dpt}{\ensuremath{\Delta{\it p}_{\rm T}}\xspace}
\newcommand{\dptmean}{\ensuremath{\langle\Delta{\it p}_{\rm T}\rangle}\xspace}
\newcommand{\Rjet}[1]{{\ensuremath{{\it R}_{\rm jet}=}   #1}\xspace}

\newcommand{\cmnt}[1]{}

\newcommand{\jetptrange}[2]{{#1 $<$ \pTchjet $<$ #2  \GeVc}\xspace}
\newcommand{\ptjet}{\ensuremath{{\it p}_{\rm T}^{\rm jet}}\xspace}
\begin{document}%

%%%%%%%%%%%%%%%  Title page %%%%%%%%%%%%%%%%%%%%%%%%
\begin{titlepage}
\PHyear{2026}
\PHnumber{175}      % required, will be obtained from CERN
\PHdate{12 June}  % required, will be obtained from CERN
%

%%% Put your own title + short title here:
\title{Modification of jet-energy flow in heavy-ion collisions} 

\ShortTitle{Jet-energy flow in heavy-ion collisions} % appears on right page headers

%%% Do not change the next lines
\Collaboration{ALICE Collaboration\thanks{See Appendix~\ref{app:collab} for the list of collaboration members}}
\ShortAuthor{ALICE Collaboration} % appears on left page headers, do not change

\begin{abstract}
The ALICE Collaboration presents the first measurements of the jet-energy flow (\dpt) observable in proton-proton and heavy-ion collisions. Jets are excellent probes for the quark--gluon plasma, a deconfined state of matter produced in heavy-ion collisions. The jet-energy flow observable characterizes the radial distribution of energy from the jet axis in an infrared and collinear-safe way and is sensitive to medium-induced parton-shower modifications. Inclusive charged jets are measured in Pb--Pb collisions at $\sqrtsNN = 5.02$~TeV for the transverse-momentum interval $60$--$80~\mathrm{GeV}/c$. For pp collisions at $\sqrt{s}=13~\TeV$, measurements include the $40$--$60$ and $60$--$80~\mathrm{GeV}/c$ intervals, where the latter serves as the reference for investigating medium-induced modifications. Results show that most parton energy is concentrated in the jet core, with a clear suppression of energy flow in heavy-ion collisions at larger radii (significance 3.5--4.5$\sigma$) indicating a narrowing of the energy flow. While all models -- PYTHIA 8, HERWIG, JEWEL, and JETSCAPE -- reproduce the pp results with only small deviations in the tails, the relative modification in Pb--Pb collisions is well described by JEWEL without recoil. Conversely, JEWEL with recoil (medium response) and JETSCAPE show significant deviations, exhibiting increasing or more constant trends with radius that are disfavored by the data.

\end{abstract}
\end{titlepage}
\setcounter{page}{2}

\section{Introduction}
High-energy collisions at the Large Hadron Collider (LHC) provide an ideal environment to study strongly interacting matter and to test the theory of quantum chromodynamics (QCD). In particular, heavy-ion collisions create extreme conditions of temperature and energy density, under which ordinary hadronic matter undergoes a transition to a deconfined state of quarks and gluons, known as the quark–gluon plasma (QGP)~\cite{Bjorken:1982qr, Busza:2018rrf}. Jets, which are collimated sprays of particles, serve as multi-scale probes of the QGP. Originating from hard scatterings of the initially interacting partons, their production cross sections can be calculated within perturbative QCD, making them quantitatively well-controlled observables.

As these partons propagate through the QGP, they lose energy to the medium and their internal structure is modified, an effect known as jet quenching~\cite{Bjorken:1982tu, PhysRevD.33.717, PhysLettB.243, Kogler_2019, ATLAS:2010isq}. The aforementioned modifications to the jet and its substructure arise from the interplay between medium-induced emissions, which redistribute energy to larger angles, combined with the stronger suppression of wide jets due to incoherent energy loss of resolved subjets (color decoherence) ~\cite{Mehtar-Tani:2011hma, Milhano:2015mng, PhysRevLett.116.211603,Chesler:2015nqz, CMS:2024zjn}. By investigating the mechanisms of jet energy loss, one gains direct insight into these parton–medium interactions and the transport properties of the QGP~\cite{ALICE:2023dwg, ALICE:2023waz, Acharya_2025}.

A number of jet and jet-substructure measurements have been previously reported by ALICE~\cite{ALICE:2022mur, ALICE:2023waz, Acharya_2025}, ATLAS~\cite{ATLAS:2022zbu, ATLAS:2010isq, ATLAS:2022vii} and CMS~\cite{CMS:2011iwn, Sirunyan:2019rfa, CMS:2019jis, CMS:2021vui} to study these parton-medium interactions. In this paper, we explore a new observable, known as the jet-energy flow (JEF) \dpt, which provides an event-by-event measure of these modifications. Defined in Eq.~\ref{Eq:Dpt-definition}, it is an infrared- and collinear-safe observable that characterizes how the energy inside a reconstructed jet is distributed as a function of the angular opening (radius) from the jet axis. Consequently, \dpt is directly sensitive to medium-induced modifications of parton-shower evolution.
\begin{equation}
\Delta p_{\rm T} = p_{{\rm T},R_{\rm i+1}} - p_{{\rm T},R_{\rm i}},
\label{Eq:Dpt-definition}
\end{equation}
where $p_{{\rm T},R_{\rm i}}$ denotes the transverse momentum of the jet reconstructed with the smaller radius and $p_{{\rm T},R_{\rm i+1}}$ that of the matched jet reconstructed with the larger radius within a similar $\eta$–$\varphi$ region in the same event. Here, $\eta$ and $\varphi$ are the pseudorapidity and azimuthal angle, respectively; further details regarding the matching procedure are provided in the section~\ref{3.1-Jet_algorithm}. The difference between these two values probes the radial distribution of jet energy relative to the jet axis.

The differential jet shape observable which measures the radial energy-density profile around the jet axis in annular regions is closely related to the \dpt observable~\cite{CMS:2018zze}. The main difference between the two observables is that the axes of the jets with different radii in the energy flow measure are not necessarily precisely aligned, while the radial profile is measured with respect to a fixed axis. In addition, the energy-flow measurement provides an event-by-event characterization of parton-shower modification induced by the QGP.

In this paper, the ALICE Collaboration presents the first measurements of the JEF \dpt in proton-proton (pp) and heavy-ion collisions. The measurement is performed for inclusive (a mixture of gluon- and quark-initiated jets) charged jets in Pb--Pb collisions at $\sqrt{s_{\mathrm{NN}}}=5.02$ TeV in the transverse-momentum interval $60$–$80~\mathrm{GeV}/c$. In pp collisions, measurements are carried out in the intervals $40$–$60~\mathrm{GeV}/c$ and $60$–$80~\mathrm{GeV}/c$ at $\sqrt{s}=13~\TeV$, with the latter used as the reference for investigating medium-induced modifications. To maximize the available statistics, pp collisions at $\sqrt{s}=13~\TeV$ (integrated luminosity $\mathcal{L}_{\rm int}$ = $25.81\pm0.43$ nb$^{-1}$~\cite{ALICE:2022mur,ppXsec}) are used instead of $\sqrt{s}=5.02~\TeV$ ($\mathcal{L}_{\rm int}$ = $18.0\pm0.4$ nb$^{-1}$~\cite{ALICE:2022mur,ppXsec}) for the reference measurement. The JEF modification, defined as the ratio of the measured JEF in Pb--Pb collisions at 5.02 TeV to that in pp collisions at 13 TeV, is compared to a PYTHIA-based ratio of pp collisions at 5 TeV to pp collisions at 13 TeV, which accounts for the beam-energy dependence of the pp baseline. The PYTHIA-based baseline is used for comparison because it provides the best description of the pp measurements, compared to other models. The results are also compared to predictions from Monte-Carlo (MC) event generators and theoretical models, including the JETSCAPE and JEWEL model, providing insight into QCD dynamics in pp and Pb--Pb collisions. These measurements place new constraints on models of jet–medium interactions and demonstrate the sensitivity of the jet-energy flow observables to the microscopic mechanisms of parton energy loss in the QGP.
 
\section{Experimental setup and data sets}
\label{sec:exp_data}
This analysis uses charged-particle tracks reconstructed with the ALICE central-barrel tracking system, covering $|\eta|<0.9$ inside a solenoidal magnet. The detectors used are the Inner Tracking System (ITS)~\cite{ALICE:2010tia} and the Time Projection Chamber (TPC)~\cite{Alme:2010ke}. This system provides high tracking efficiency in the range $0.15 < \pt < 100~\mathrm{GeV}/c$, with uniform acceptance in $|\eta|<0.9$ and $0 < \varphi < 2\pi$.  The ITS is the innermost tracking detector of ALICE, designed for high-resolution tracking and vertexing. It consists of six cylindrical layers of silicon detectors: two layers each of Silicon Pixel Detectors (SPD), Silicon Drift Detectors, and Silicon Strip Detectors. The ITS  enables precise reconstruction of primary and secondary vertices and particle tracking at low transverse momentum. The TPC is the main tracking detector in ALICE and it plays a crucial role in particle identification. It is a large cylindrical gas-filled chamber with readout planes at both ends. Charged particles traversing the TPC ionize the gas, and the resulting electrons drift toward the readout planes under an applied electric field.  A detailed description of the ALICE detector and its performance is provided in Refs.~\cite{ALICE:2014sbx,ALICE:2008ngc}.

The results presented in this paper are based on pp collisions at a centre-of-mass energy of $\sqrt{s}=13~\TeV$ recorded by ALICE during 2016--2018 and Pb--Pb collisions at $\sqrtsNN=5.02~\TeV$ recorded in 2018. The MC samples used for the corrections described in Section~\ref{sec:corrections} are generated using the PYTHIA 8.2 event generator~\cite{Sjostrand:2014zea} with the Monash 2013 tune~\cite{Skands:2014pea}, and with the GEANT 3.21.11~\cite{Brun:118715} transport model. The ALICE detector layout, along with variations in the data-taking conditions during the run, are reproduced in the simulation.

\section{Analysis method}\label{sec:analysis}
Events used in this analysis are selected using the minimum bias (MB) trigger, defined by coincident hits on the V0 scintillator detectors~\cite{ALICE:2013axi} and a time selection on the coincidence of the signals in the Zero-Degree Calorimeters (ZDCs)~\cite{Aamodt:2008zz} on both sides of the experiment. The V0 detector consists of two scintillator arrays placed on either side of the interaction point along the beam axis, covering the pseudorapidity ranges $2.8 < \eta < 5.1$ (V0A) and $-3.7 < \eta < -1.7$ (V0C). The MB trigger is activated by a simultaneous signal in V0A and V0C, coinciding with a bunch crossing. When ZDC information is available, a further selection is made on the timing of the ZDC coincidence compared to the bunch-crossing time to reject out-of-time background. For Pb--Pb data, in addition to the MB trigger, an online centrality trigger is applied to select the most central events. This analysis focuses on the 0--10\% centrality class, where jet quenching effects are expected to be strongest. To ensure uniform detector acceptance, events are required to have a primary vertex located within $\pm$10 cm of the nominal interaction point along the beam direction. Events with more than one reconstructed vertex are excluded to avoid pileup effects. 

ALICE achieves high reconstruction efficiency for charged particles within the transverse-momentum range of $0.15 < \pt <100 \ \GeVc$. The single track reconstruction efficiency rises from $\sim 60\%$ at 0.15~\GeVc, plateauing at $\sim 80\%$ for tracks with  $p_{\rm T} >$ 0.4~\GeVc. The track transverse-momentum resolution, $\sigma_{p_{\rm T}}/p_{\rm T}$, is approximately $ 1\%$ at {\mbox{$\pt=1$ \GeVc}}, increasing to about $ 4\%$ at $\pt=50$ \GeVc. 
 
To ensure uniform coverage in the $\eta$--$\varphi$ plane, the analysis utilizes a ``hybrid track'' sample consisting of three categories. First, global tracks must include at least one hit in the SPD and satisfy quality criteria~\cite{ALICE:2019wqv}, including $\ge 70$ TPC space points (out of a maximum 159) and $\ge 80\%$ of the geometrically findable clusters. Second, complementary tracks lack SPD hits but meet all other standard criteria. Third, the sample includes tracks with a poor fit, characterized by a high $\chi^{2}$ value or missing expected hits in the ITS layers. To ensure momentum accuracy and reduce contamination, both complementary and poor-fit tracks are refitted with a constraint to the primary vertex of the event. This analysis selects tracks with $p_{\text{T}} \ge 0.15~\text{GeV}/c$ within $|\eta| \le 0.9$ over the full azimuth ($0 < \varphi < 2\pi$).

\subsection{Jet reconstruction and jet-energy flow calculation}
\label{3.1-Jet_algorithm}
The selected tracks serve as input for the jet-clustering algorithms. In this analysis, jets with varying radii from $R =$ 0.05 to 0.4, in steps of $\Delta R_{\rm step}=$ 0.05 (i.e., $R =$ 0.05, 0.1, 0.15, 0.2, 0.25, 0.3, 0.35, and 0.4), are reconstructed in the same event to examine the \pt difference between jets. Apart from the jet-resolution parameter, the same configuration is used for the reconstruction of all jets. Jets are reconstructed using the anti-$k_{\rm T}$ clustering algorithm~\cite{Cacciari_2008}, as implemented in FastJet~\cite{Cacciari:2011ma}. Charged jets are considered and clustered using the $E$-recombination scheme, in which the four-momenta of the jet constituents are combined. The reconstructed jets are required to satisfy $|\eta_{\rm jet}| \leq 0.5$ and a minimum jet \pt of $10~\GeVc$. A ghost area of 0.005 is used for the jet-area determination, where low-momentum ghost particles are added uniformly in the event to estimate the active jet area. In Pb--Pb collisions, jets are additionally required to contain a leading hadron with $\pt \geq 8~\GeVc$.

\begin{figure}[hbt!]
\centering
\includegraphics[scale=0.3]{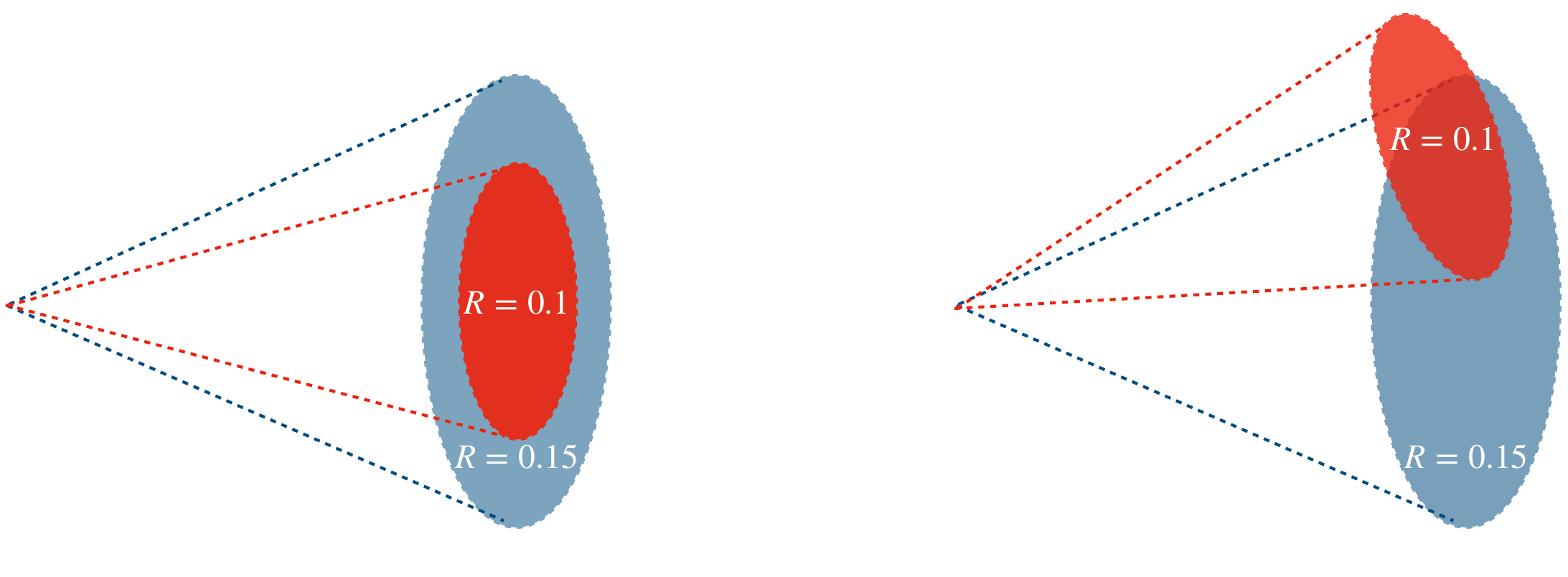} 
\caption{Sketch of the matching process between jets of different resolution parameter. On the left, a case where the jet axes are completely aligned and on the right a more general case where there is significant overlap between the jets but the axes themselves are not aligned.}
\label{fig:Jet-matching-sketch}
\end{figure}

Jets are then paired into ``[narrower, wider]'' combinations to ensure sufficient area overlap, using adjacent radius pairs: [$R_{\mathrm {jet_1}}, R_{\mathrm {jet_2}}$] $=$ [0.05, 0.1], [0.1, 0.15], [0.15, 0.2], [0.2, 0.25], [0.25, 0.3], [0.3, 0.35], [0.35, 0.4]. To select true [$R_{\mathrm {jet_1}}, R_{\mathrm {jet_2}}$] matched pairs, bijective matching is performed within $\Delta R < 0.2$. In the bijective matching procedure, each $R_{\mathrm {jet_1}}$ jet is matched to its closest $R_{\mathrm {jet_2}}$ jet, and vice versa. If two jets identify each other as their respective closest match, they form a valid jet pair. The matching criterion is based solely on the jet-axis positions and does not require a shared \pt fraction between the jets. In practice, the axes of matched jets with different radii are found to be strongly aligned, typically with $\Delta R < 0.05$. For each valid jet pair, the energy flow is probed by calculating the transverse momentum difference \dpt as $\dpt = p_{\rm T,2} - p_{\rm T,1}$, where $p_{\rm T,1}$ and $p_{\rm T,2}$ denote the transverse momenta of jet$_1$ and jet$_2$, respectively, with jet$_2$ having the larger radius ($R_{\mathrm {jet_2}} > R_{\mathrm {jet_1}}$). This procedure is repeated until all jets in each radius pair are exhausted. A schematic illustration of the matching procedure between jets of different radii is shown in Fig.~\ref{fig:Jet-matching-sketch}.  The 
\dpt observable is evaluated in the \jetptrange{40}{60} and \jetptrange{60}{80} intervals in pp collisions, and in the \jetptrange{60}{80} interval in Pb–Pb collisions, where $p^{\mathrm{ch.\ jet}}_{\mathrm T}$ refers to $p_{\rm T,1}$, i.e., the transverse momentum of the smaller-radius jet.
 
\subsection{Corrections}\label{sec:corrections}
\subsubsection{Background subtraction}
\label{3.3-Bkg_subtraction}
In pp collisions, background contributions from the underlying event (UE) are negligible in the studied jet-\pt interval. Furthermore, any small residual contribution is further reduced since we examine the difference in \pt between jets that lie in close proximity within the $\eta$–$\varphi$ plane. In Pb--Pb collisions, a substantial background of uncorrelated soft particles originating mostly from soft production processes is present. This background can significantly inflate the measured jet energy and distort its internal structure. The thermal background is subtracted using the event-by-event constituent-subtraction method~\cite{Berta:2014eza}. This subtraction method introduces "ghost particles" to the event with transverse momentum $p_{\rm T}^g = \rho \cdot A_g$, where $\rho$ is the median background density and $A_g$ is the small area occupied by a single ghost. The method iteratively identifies the closest pair of a real particle and a ghost particle in the $\eta-\varphi$ plane. For each identified real-ghost particle pair, provided their separation is within a maximum distance, $\Delta R_{\text{max}}$ (a key free parameter of the method), the momentum of the softer of the two particles is subtracted from the harder one. The four-momentum of the softer particle is then set to zero. This step is repeated for all matched pairs. After the iteration is complete, all particles (real or ghost) with zero momentum are discarded from the event. The jet-finding algorithm is then applied to the resulting collection of background-subtracted particles to reconstruct the background-corrected jets. It should also be noted that due to the definition of the energy flow observable as \dpt, the contributions of shared background fluctuations in the matched jets will cancel.

The free parameter $\Delta R_{\text{max}}$ controls the locality of the subtraction. In this analysis, its value is tuned based on the jet radius being studied to avoid over- or under-subtraction. For jets with $R \le 0.2$, a value of $\Delta R_{\text{max}} = 0.1$ is used, while for larger jets, $\Delta R_{\text{max}} = 0.25$ is employed. The sensitivity to this choice is evaluated as a systematic uncertainty.

\subsubsection{Unfolding procedure}
\label{3.4-Unfolding}
The background-subtraction procedure corrects only for the average UE, while the remaining background fluctuations and detector effects are corrected using an unfolding procedure. In this procedure, the correspondence between the true (particle-level) and measured (detector-level) distributions, encoded in the response matrix (RM), is established using MC simulations through a jet-matching procedure. A four-dimensional RM $(\dpt^{\rm det},\dpt^{\rm truth},p^{\rm ch.\ jet}_{\rm T,det},p^{\rm ch.\ jet}_{\rm T,truth})$, describing the detector and background response in \ptjet and \dpt, is constructed from jets matched between detector-level (combined-level) and particle-level, and is used in a two-dimensional unfolding with the iterative Bayesian algorithm~\cite{bayesianUnfolding} as implemented in RooUnfold~\cite{RooUnfold}. 

The pp MC sample used to construct the RM consists of events simulated with the PYTHIA~8 event generator (Monash2013 tune~\cite{Skands:2014pea}) and passed through the ALICE detector simulation using GEANT~3 to account for particle interactions with the detector material. In the Pb--Pb case, PYTHIA~6 events including GEANT~3 detector response (detector-level) are embedded into reconstructed Pb--Pb data events (combined-level). Background subtraction and jet reconstruction are performed on these combined-level events in the same manner as in the data analysis. Jets are then matched geometrically between the detector-level (in pp) or combined-level (in Pb--Pb) and the corresponding particle-level (truth-level) jets, with the additional requirement that the matches be unique. The kinematic ranges utilized for jets are $0 < p_{\text{T,truth}}^{\text{ch. jet}} < 150~\text{GeV}/c$ and $10 < p_{\text{T,det}}^{\text{ch. jet}} < 140~\text{GeV}/c$.

The unfolding convergence test is performed to optimize the number of unfolding iterations, and the optimal value is found to be four for both pp and Pb--Pb collisions. Closure tests are also performed to ensure the robustness and reliability of the unfolding procedure. Two types of closure tests are carried out. A refolding test is performed to verify that the detector-level spectrum can be recovered by reversing the unfolding procedure. In this test, the unfolded result is multiplied by the response matrix, and the resulting spectrum is compared with the detector-level spectrum. In addition, a closure test is performed in which simulated data at the detector level are unfolded and compared with the corresponding truth-level distributions. For the test, the simulated data are divided into two statistically independent subsets: one is used to determine the response matrix, while the other serves as a test sample. For both closure tests, good agreement is observed, with only small bin-by-bin fluctuations.
 
\section{Systematic uncertainties}
\label{sec:systematics}
The kinematic selections and background correction methods have inherent limitations and associated uncertainties. To assess the resilience of the final measurement, the relevant analysis parameters are varied within the limits of their known uncertainties, and the resulting impact on the final result is evaluated. The total uncertainty is calculated as the quadratic sum of the individual contributions listed below assuming that these contributions are uncorrelated.
\begin{enumerate}
    \item \textbf{Tracking Efficiency:} This is a dominant source of uncertainty, arising from the imperfect knowledge of the detector's efficiency in reconstructing charged-particle tracks.
    The tracking efficiency uncertainty is estimated to be 3\% in pp collisions and 3--5\% in Pb–Pb collisions, depending on track \pt,  based on variations of the track selection criteria and the uncertainty in the ITS–TPC track-matching efficiency in the MC simulation ~\cite{Acharya_2025}. The associated systematic uncertainty is evaluated by creating a modified response matrix where a fraction of tracks equal to the corresponding uncertainty are randomly rejected. The final unfolded data using this modified response is compared to the default result to determine the systematic uncertainty.
    \item \textbf{Unfolding Procedure}
This addresses uncertainties inherent in the statistical unfolding method itself. It includes three variations:
\begin{itemize}
    \item[--] \textbf{Regularization parameter ($N_{\text{iter}}$):} The number of iterations in the Bayesian unfolding algorithm is varied by $\pm2$ (number based on the performance of the closure tests) around the nominal value to test the stability of the result. The average difference between the results obtained with these two variations is selected as the relative systematic uncertainty. 
    \item \textbf{Truncation:} The lower limit of the jet-\pt range used at the detector level is varied by $\pm5$ \GeVc~ to check for sensitivity to the chosen kinematic boundaries.
    \item \textbf{Prior variation:} The dependence on the shape of the true distribution from MC (the prior) is tested. In order to evaluate this systematic, we fit the prior \dpt distribution with an exponential function. The slope of this function is varied by $\pm5\%$ and a new response is constructed for each variation by scaling the prior with \dpt-dependent weight. The raw distribution is then unfolded using the modified response matrices. The average deviation of the resulting ratios with respect to the default result is assigned as the relative systematic uncertainty associated with this source. 
\end{itemize}
These three sources probe the same source of uncertainty, so the total unfolding uncertainty is estimated here as the standard deviation of all three contributions from zero.
\item \textbf{Model Dependence:}
This uncertainty accounts for the choice of the MC event generator used to compute the corrections and the response matrix.
\begin{itemize}
    \item \textbf{pp}: To investigate the sensitivity of the final results to the choice of event generator used for deriving the correction factors, a comparison of unfolded results obtained with different generators is performed. To reduce the computing cost for this test, the study uses a fast simulation with parametrized response of the ALICE detector instead of a full simulation using GEANT. Simulation samples were generated with both PYTHIA and HERWIG~\cite{Bellm:2015jjp} and passed through this parametrized detector response. The corresponding response matrices were then constructed and used to unfold the raw-data distribution. The relative difference between the results unfolded with HERWIG and those unfolded with PYTHIA is taken as the systematic uncertainty due to model dependence.
    \item \textbf{Pb--Pb}: The same procedure is applied for Pb--Pb collisions, using the JEWEL~\cite{Zapp:2013vla} generator without the inclusion of medium partons recoiling after interactions with the jet to evaluate the model-dependence systematic uncertainty.
\end{itemize}
\item \textbf{Background Subtraction Method (Pb--Pb only):}
This uncertainty is specific to the heavy-ion analysis and quantifies the sensitivity to the parameters of the constituent subtraction method. The free parameter $\Delta R_{\text{max}}$ is varied to explore potential over- or under-subtraction of the background. To evaluate the systematic uncertainty, $\Delta R_{\text{max}}$ is varied to $0.025$ for $R \leq 0.2$ and to $0.3$ for larger jet radii. The full analysis is repeated for data and embedded MC with the varied parameter, and the difference in the final unfolded result is taken as the systematic uncertainty.
\item \textbf{Leading Track \pt Cut (Pb--Pb only):}
This uncertainty assesses the potential bias introduced by the requirement that jets in Pb--Pb collisions contain a high-\pt leading track, which favors jets with hard fragmentation. The analysis is repeated for both data and embedded MC using a stricter cut of $p_{\rm T,\text{leading}} \ge 10$ \GeVc~(compared to the default of 8 GeV/c). The relative difference in the final unfolded distributions is assigned as the systematic uncertainty.
\end{enumerate}

The total systematic uncertainty is calculated by assuming that the individual sources are uncorrelated and by summing in quadrature the contributions from the unfolding procedure, tracking efficiency, and model dependence. A summary of the systematic uncertainties is provided in Tables~\ref{Table:Systematics_pp_40_60},~\ref{Table:Systematics_pp_60_80}, and~\ref{Table:Systematics_PbPb_60_80}.

\begin{table}[h!]
\centering
\caption{Systematic uncertainties in \jetptrange{40}{60} interval in pp collisions at $\sqrt{s}= 13$~ TeV. The range denotes the maximum and minimum uncertainties across the \dpt intervals.}
\label{Table:Systematics_pp_40_60}
\begin{tabular}{|l|c|c|c|c|c|c|c|}
\hline
\diagbox[width=3.5cm, height=1cm]{
Source}{ $[R_{\rm {jet_{1}}}, R_{\rm {jet_{2}}}]$} & [0.05, 0.1] & [0.1, 0.15] & [0.15, 0.2] & [0.2, 0.25] & [0.25, 0.3] & [0.3, 0.35] & [0.35, 0.4] \\
\hline
Tracking & 2--4\% & 1--4\% & 1--3\% & 1--2\% & 2--3\% & 1--3\% & 1--3\% \\
\hline
Model & 1--3\% & 1--2\% & 1--5\% & 1--4\% & 2--4\% & 2--3\% & 3--4\% \\
\hline
$N_{\text{iter}}$ & 0--1\% & 0--2\% & 0--1\% & 0--1\% & 0--1\% & 0--1\% & 0--1\% \\
\hline
Prior  & 0--1\% & 0--1\% & 0--1\% & 0--1\% & 0--2\% & 0--4\% & 0--3\% \\
\hline
Truncation & 0--1\% & 0--1\% & 0--1\% & 0--1\% & 0--1\% & 0--1\% & 0--1\% \\

\hline
\hline
Total & 2--5\% & 1--5\% & 1--6\% & 1--5\% & 2--5\% & 1--5\% & 3--5\% \\
\hline
\end{tabular}
\end{table}

\begin{table}[h!]
\centering
\caption{Systematic uncertainties in \jetptrange{60}{80} interval in pp collisions at $\sqrt{s}=13$~ TeV. The range denotes the maximum and minimum uncertainties across the \dpt intervals.}
\label{Table:Systematics_pp_60_80}
\begin{tabular}{|l|c|c|c|c|c|c|c|}
\hline
\diagbox[width=3.5cm, height=1cm]{
Source}{ $[R_{\rm {jet_{1}}}, R_{\rm {jet_{2}}}]$} & [0.05, 0.1] & [0.1, 0.15] & [0.15, 0.2] & [0.2, 0.25] & [0.25, 0.3] & [0.3, 0.35] & [0.35, 0.4] \\
\hline
Tracking & 3--4\% & 0--3\% & 0--5\% & 1--6\% & 1--4\% & 2--3\% & 1--4\% \\
\hline
Model & 1--3\% & 1--4\% & 1--3\% & 2--5\% & 3--5\% & 2--10\% & 3--10\% \\
\hline
$N_{\text{iter}}$ & 0--2\% & 1--2\% & 0--1\% & 0--4\% & 1--4\% & 1--5\% & 1--5\% \\
\hline
Prior  & 0--1\% & 0--1\% & 0--2\% & 0--2\% & 0--2\% & 0--2\% & 0--2\% \\
\hline
Truncation & 0--1\% & 0--1\% & 0--1\% & 0--1\% & 0--1\% & 0--1\% & 0--1\% \\
\hline
\hline
Total & 3--5\% & 1--5\% & 1--6\% & 2--8\% & 3--7\% & 3--11\% & 3--11\% \\
\hline
\end{tabular}
\end{table}

\begin{table}[h!]
\centering
\caption{Systematic uncertainties in \jetptrange{60}{80} interval in Pb--Pb collisions at $\sqrt{s_{\mathrm {NN}}}= 5.02$~ TeV. The range denotes the maximum and minimum uncertainties across the \dpt intervals.}
\label{Table:Systematics_PbPb_60_80}
\begin{tabular}{|l|c|c|c|c|}
\hline
\diagbox[width=4cm, height=1cm]{
Source}{ $[R_{\rm {jet_{1}}}, R_{\rm {jet_{2}}}]$} & [0.05, 0.1] & [0.1, 0.15] & [0.15, 0.2] & [0.2, 0.25] \\
\hline
Tracking & 1--7\% & 2--9\% & 2--11\% & 2--15\%  \\
\hline
Model & 1--4\% & 2--10\% & 2--10\% & 2--10\%  \\
\hline
Background & 1--3\% & 1--5\% & 2--10\% & 2--20\%  \\
\hline
Leading track & 0--2\% & 0--2\% & 0--4\% & 0--4\%  \\
\hline
$N_{\text{iter}}$ & 0--2\% & 2--3\% & 2--7\%  & 2--18\%  \\
\hline
Prior  & 0--1\% & 0--2\% & 1--5\% & 2--22\%  \\
\hline
Truncation & 0--1\% & 0--1\% & 0--1\% & 0--1\%  \\
\hline
\hline
Total & 2--9\% & 3--15\% & 4--19\% & 4--32\%  \\
\hline
\end{tabular}
\end{table}
 
\section{Results}
\label{sec:results}
\subsection{Jet-energy flow in pp collisions}
The JEF \dpt are reported as self-normalized distributions, 
\begin{equation}
    \frac{1}{\sigma_{\rm jet\ ch,\ pair}}\frac{{\rm d} \sigma}{{\rm d} \dpt} = \frac{1}{N_{\rm jet,\ pair}}\frac{{\rm d} N}{{\rm d}\dpt},
\end{equation}
where $N_{\rm jet\, pair}$ is the number of reconstructed charged-particle jet pairs in a given event sample for a given range in $\pTchjet$, and $\sigma_{\rm jet\ ch,\ pair}$ is the corresponding cross section.  

\begin{figure}[!htb]
\centering
   \includegraphics[scale=0.39]{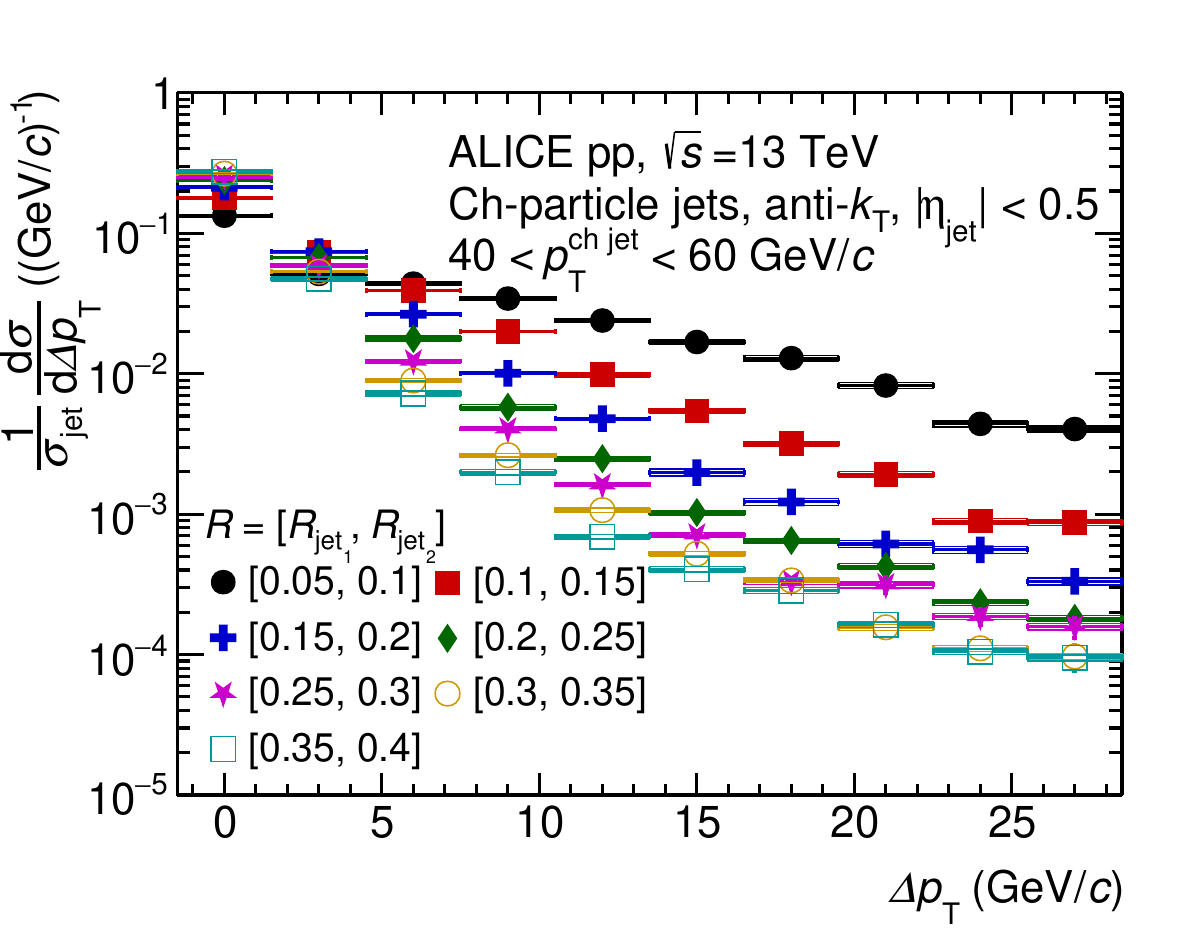}
    \includegraphics[scale=0.39]{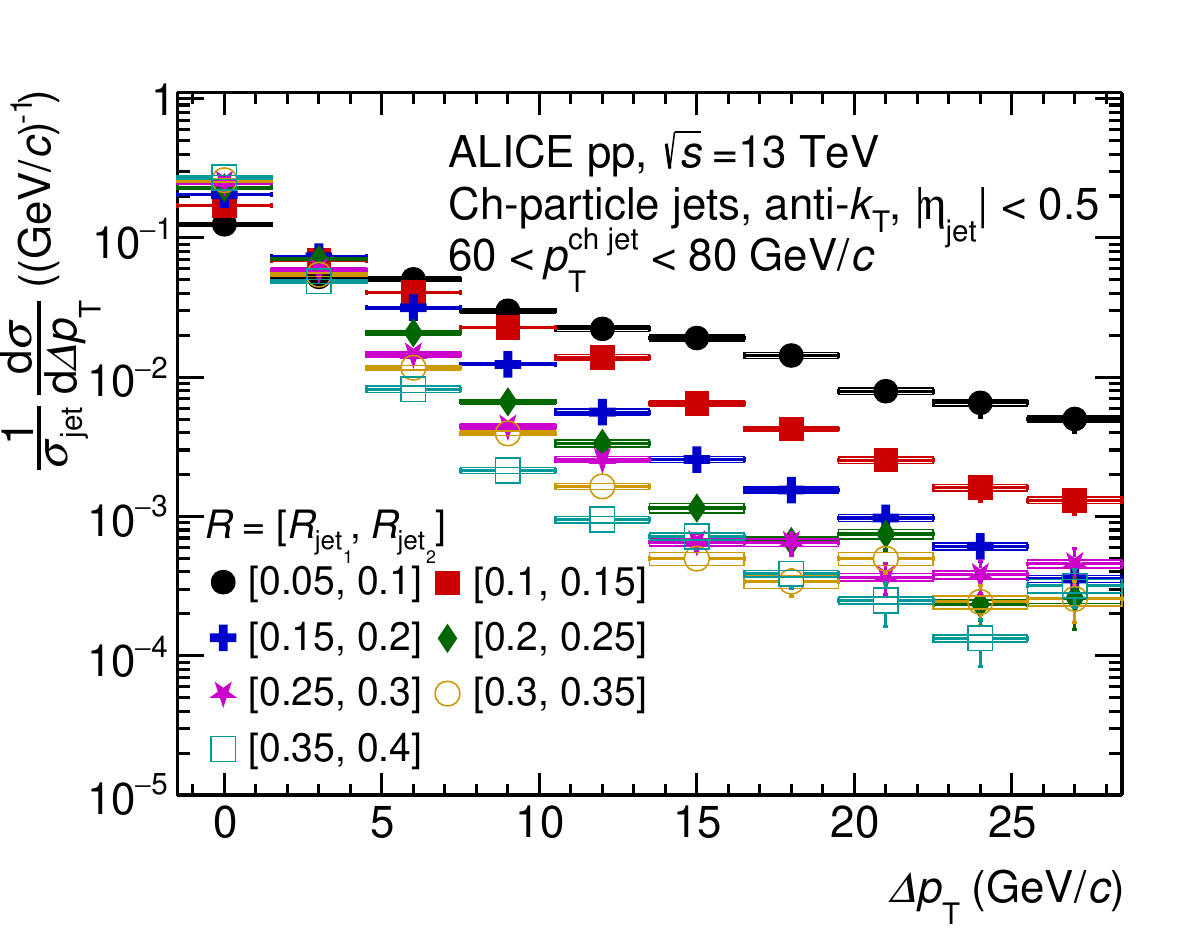}
    \caption{JEF distributions in \jetptrange{40}{60} (left) and \jetptrange{60}{80} (right) for each of the seven jet pairs with radii of [0.05, 0.1], [0.1, 0.15], [0.15, 0.2], [0.2, 0.25], [0.25, 0.3], [0.3, 0.35], and [0.35, 0.4] in pp collisions at $\sqrt{s}=13$ TeV. For certain radius-pair \dpt distributions, the statistical uncertainties are smaller than the marker size.}
\label{fig:jef_pp_data}
\end{figure}

The fully corrected JEF distributions in the intervals \jetptrange{40}{60} and {\mbox{\jetptrange{60}{80}}} for each of the seven jet pairs with radii of [0.05, 0.1], [0.1, 0.15], [0.15, 0.2], [0.2, 0.25], [0.25, 0.3], [0.3, 0.35], and [0.35, 0.4] in pp collisions at $\sqrt{s}=13$ TeV are shown in Fig.~\ref{fig:jef_pp_data}. For all cases, the systematic uncertainties are well controlled and do not exceed the statistical uncertainties. The distributions peak at $\dpt=0$ and fall steeply toward larger \dpt. The steepness is observed to increase smoothly with increasing radius, with a flatter behavior for the smallest jet radius pair, $R_{\mathrm {jet_1}}={0.05}$ and $R_{\mathrm {jet_2}}=0.1$, and the steepest fall for the largest jet radius pair, $R_{\mathrm {jet_1}}={0.35}$ and $R_{\mathrm {jet_2}}={0.40}$. This behavior reflects the fact that most of the energy flow is concentrated near the jet axis, resulting in a flatter distribution of JEF at small jet radii, which is the expected behavior of a parton shower. In addition, the distributions of the energy flow at the largest radii reported, for $R_{\mathrm {jet_1}}={0.3}$ and $R_{\mathrm {jet_2}}={0.35}$ and between $R_{\mathrm {jet_1}}={0.35}$ and $R_{\mathrm {jet_2}}={0.40}$, are found to be significantly closer to each other. A similar trend is observed for the higher jet-\pt interval.
\subsubsection{Comparison to MC models}
\begin{figure}[!htb]
\centering
   \includegraphics[scale=0.3]{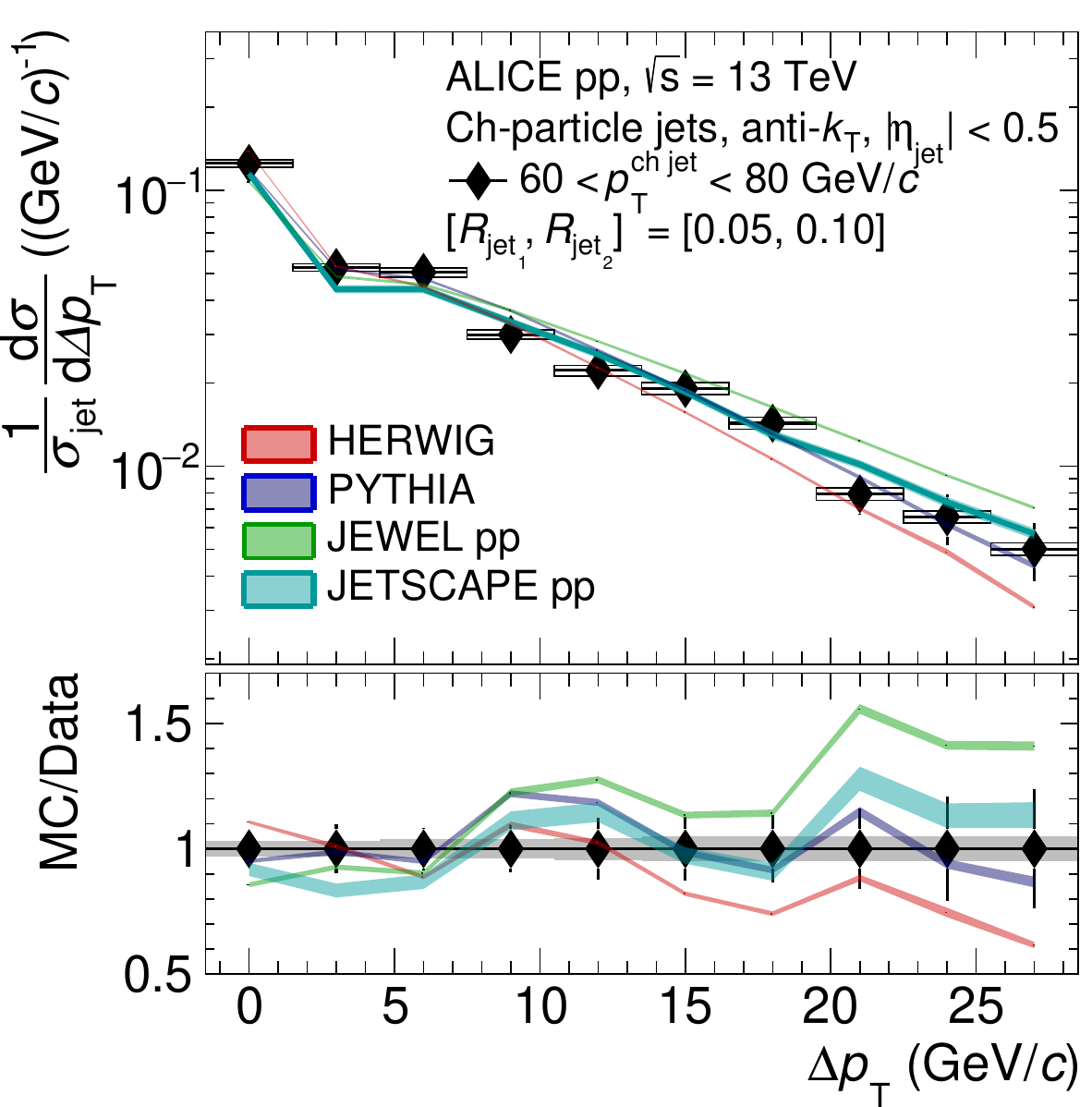}
    \includegraphics[scale=0.3]{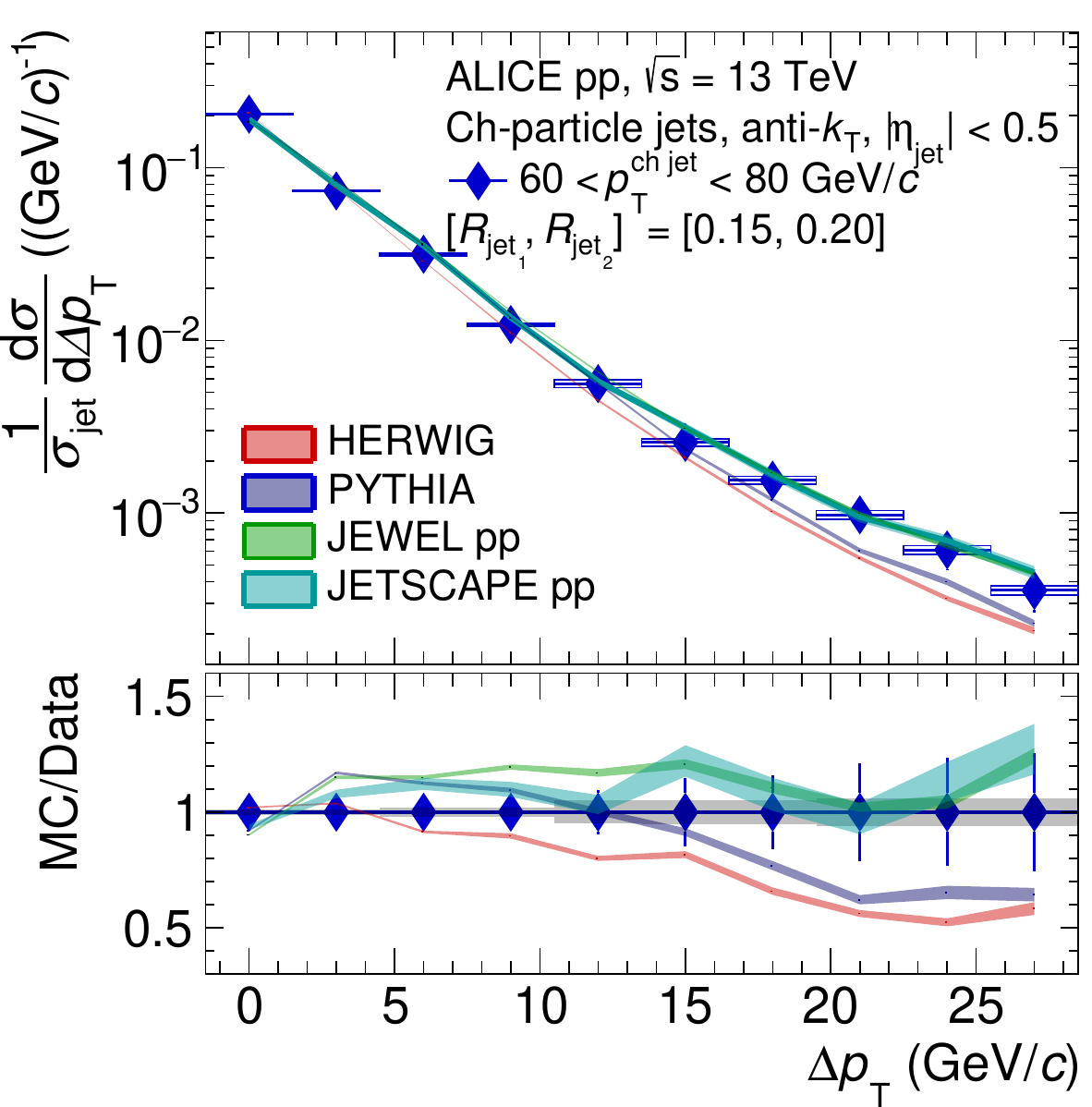}
       \includegraphics[scale=0.3]{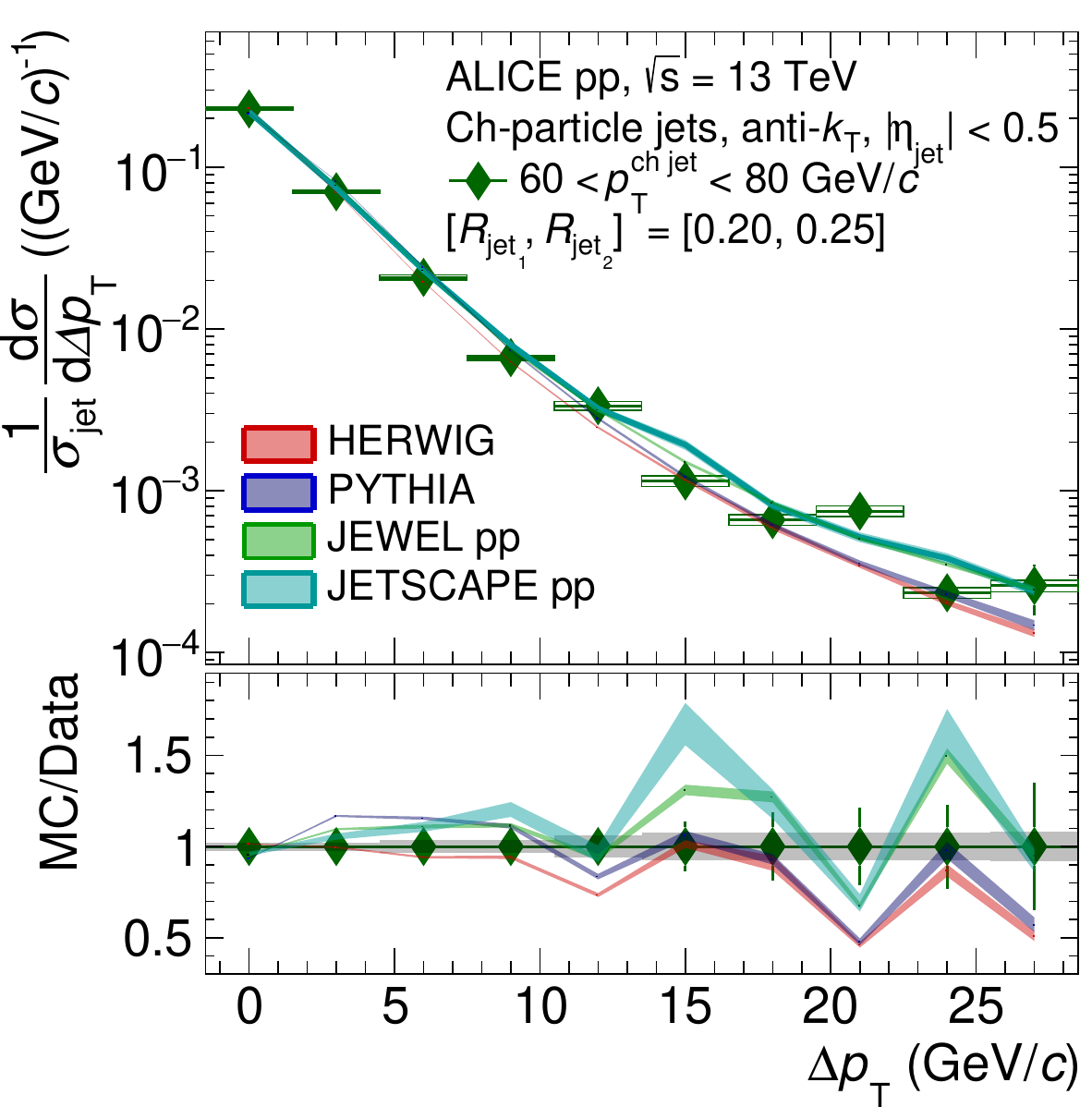}
    \caption{ JEF distribution in \jetptrange{60}{80} for three jet pairs with radii of [0.05, 0.1] (top left), [0.15, 0.20] (top right), and [0.2, 0.25] (bottom) compared to PYTHIA, HERWIG, JEWEL pp~\cite{Zapp:2013vla} and JETSCAPE (PP19 tune) ~\cite{JETSCAPE:2019udz} in pp collisions at $\sqrt{s}=13$ TeV. The bottom panels of each figure show the ratio of MC to data, and the gray band represents the systematic uncertainty of the data.}
\label{fig:jef_pp_60_80}
\end{figure}

The JEF measurement is compared to predictions from various MC event generators employing different parton-shower and hadronization schemes to assess how well the models describe the data and to investigate the role of different aspects of jet evolution, including parton showering and hadronization. Figure~\ref{fig:jef_pp_60_80} shows comparisons to PYTHIA~8 (Monash tune~\cite{Skands:2014pea}), HERWIG (default tune), JEWEL in vacuum (or pp) (default tune)~\cite{Zapp:2012ak, Zapp:2013vla}, and JETSCAPE (PP19 tune) ~\cite{JETSCAPE:2019udz}. The MC models considered here implement different parton-shower schemes, namely \pt-ordered (PYTHIA~8), and virtuality or angular-ordered (HERWIG, JEWEL). JETSCAPE uses MATTER (Modular Algorithm for Tracking Energy-loss and Radiation)~\cite{Majumder:2013re} to model the high-virtuality stage and Linear Boltzmann Transport (LBT)~\cite{He:2015pra} for the low-virtuality stage of the parton shower. These generators also employ different hadronization schemes: string breaking as described by the Lund model (PYTHIA~8, JEWEL, JETSCAPE)~\cite{Sjostrand:2014zea} and cluster fragmentation (HERWIG)~\cite{Bellm:2015jjp}. Overall, the results are well reproduced by all models, with some deviations observed in the high-\dpt tails of the distributions. The overall agreement indicates that the observable is robust against hadronization effects.

\subsubsection{Mean energy flow (\dptmean) and its jet \pt dependence}
\begin{figure}[!htb]
\centering
  \includegraphics[scale=0.505]{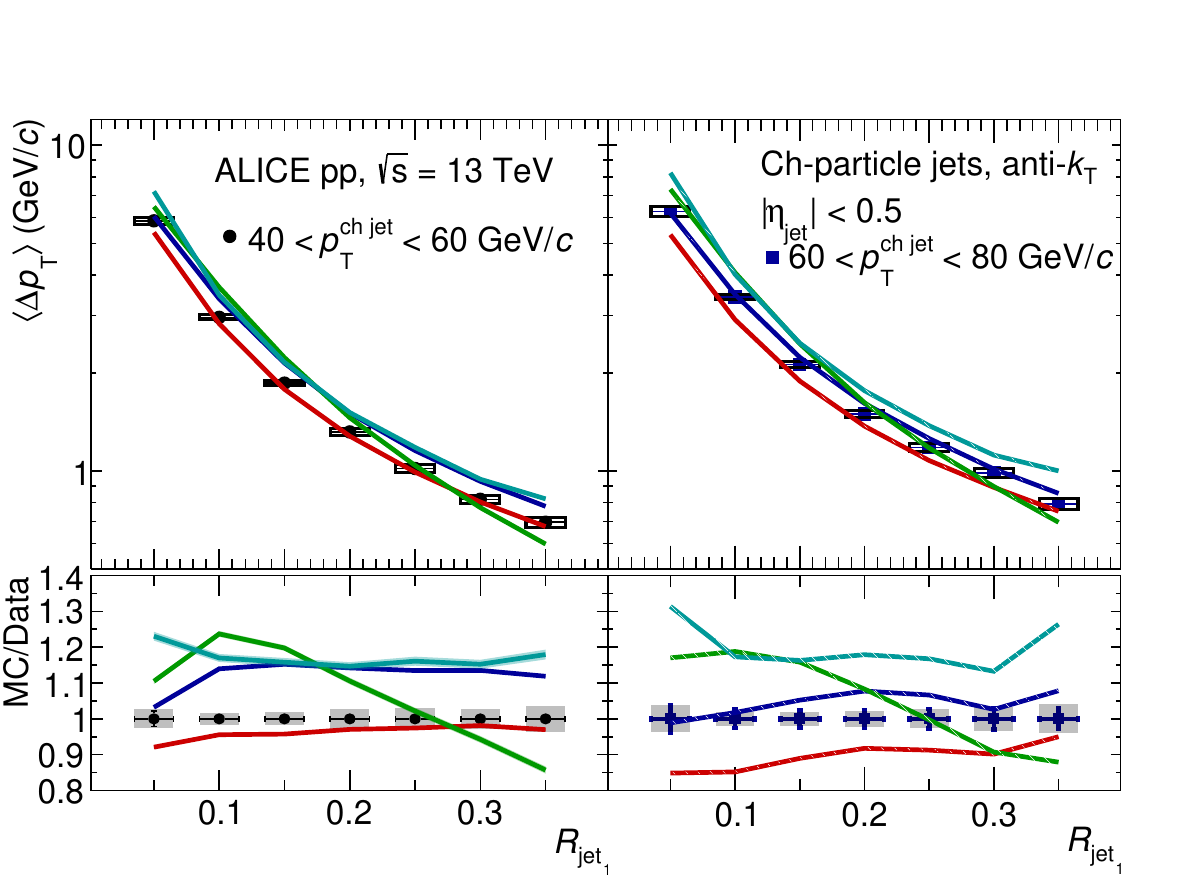}
  \hspace{-0.6cm}
    \includegraphics[scale=0.265]{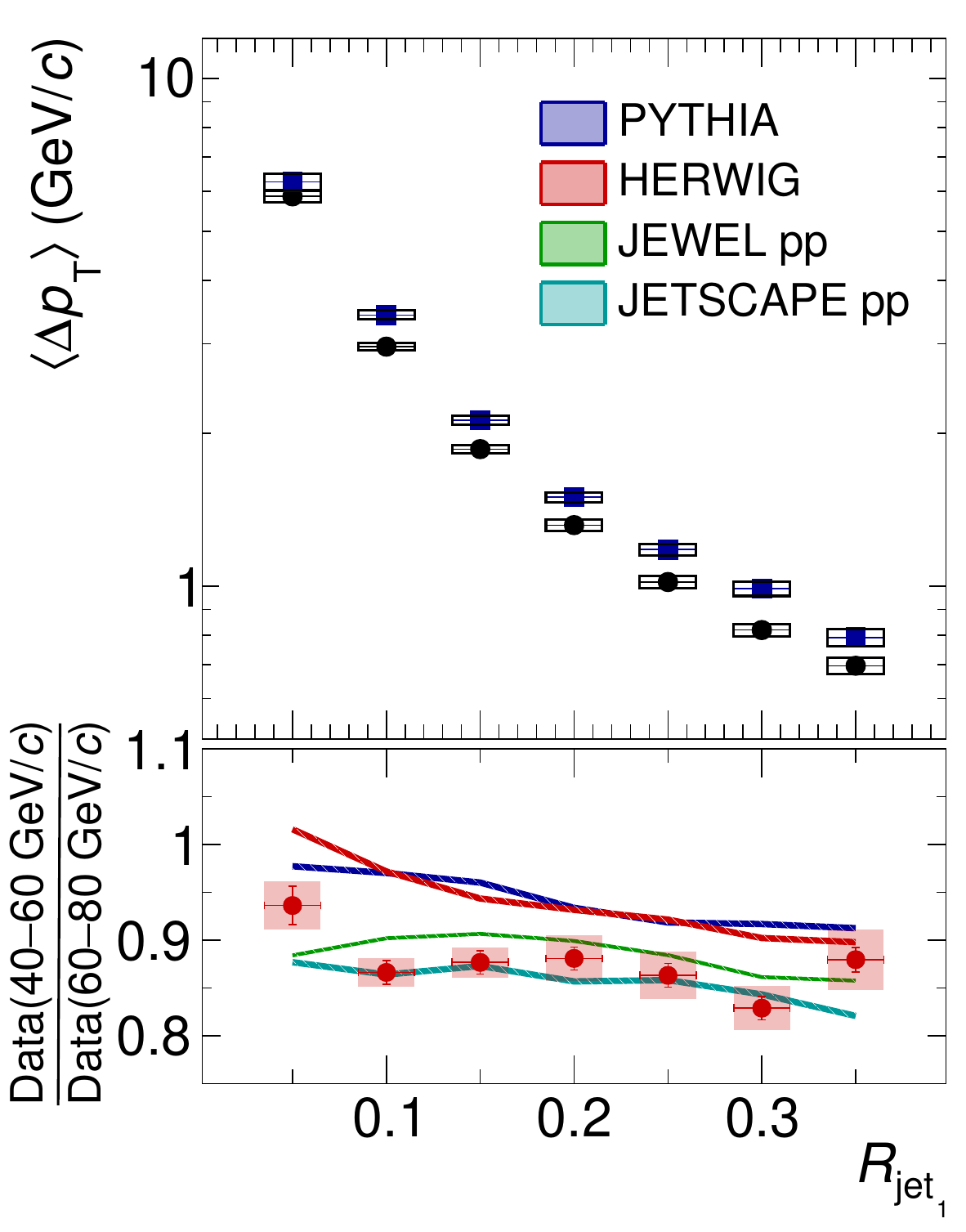}
    \caption{ Mean JEF in the \jetptrange{40}{60} (left) and \jetptrange{60}{80} (middle) jet-\pt intervals, and its jet-\pt dependence (right), as a function of the smaller jet radius, $R_{\rm jet_1}$, compared with PYTHIA, HERWIG, JEWEL, and JETSCAPE predictions for pp collisions at $\sqrt{s}=13~$TeV.}
\label{fig:jef_pp_summary}
\end{figure}

The evolution of the JEF as a function of jet radius and jet \pt is quantified by examining the mean of the \dpt distributions. Figure~\ref{fig:jef_pp_summary} summarizes the mean energy flow, \dptmean, as a function of the smaller jet radius, $R_{\rm jet_1}$, in pp collisions. The leftmost top panel shows \dptmean compared to PYTHIA~8, HERWIG, JEWEL, and JETSCAPE in the interval \jetptrange{40}{60}, the middle top panel shows the same comparison for \jetptrange{60}{80}, and the rightmost top panel shows the jet radius dependence for the two jet-\pt intervals. For both jet-\pt intervals, the mean energy flow decreases with increasing jet radius, demonstrating that the jet energy is concentrated in the core and thus reflecting the collimated nature of parton showers. The bottom panels on the left and in the middle show the ratio of data to MC predictions. HERWIG provides a better description of the mean energy flow than PYTHIA~8, JETSCAPE, and JEWEL at low jet \pt, whereas at higher jet \pt, PYTHIA~8 provides the best description. Owing to the increased statistical precision of the mean values extracted from the \dpt distributions, the ratio comparison enables a clearer discrimination between the MC models and the data. The bottom right panel shows the ratio of the mean energy flow in {\mbox{\jetptrange{40}{60}}} to that in \jetptrange{60}{80}, compared to the MC models. A weak dependence on the jet momentum is observed, which is reasonably well reproduced by the MC predictions.

\subsection{Jet-energy flow in Pb--Pb}
\begin{figure}[!htb]
\centering
   \includegraphics[scale=0.5]{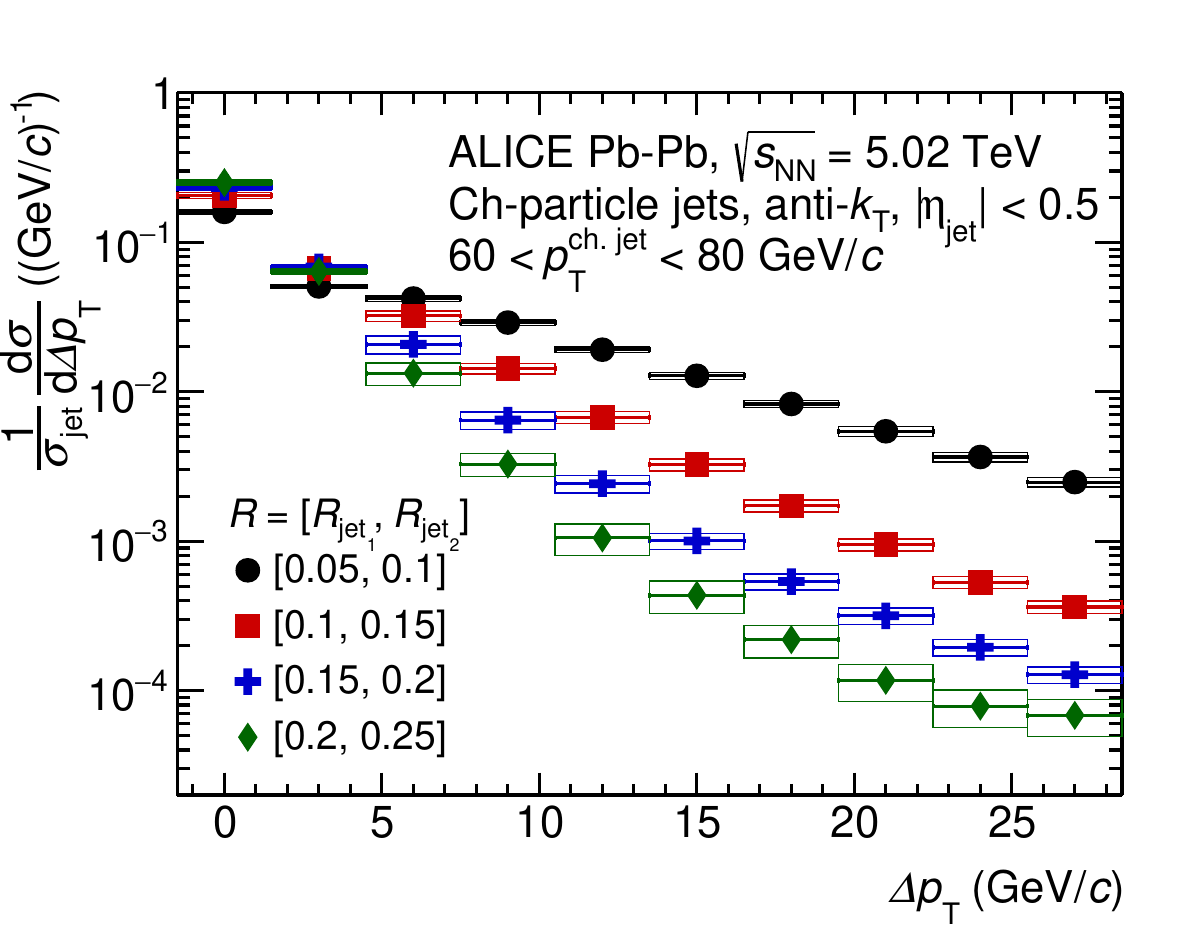}
    \caption{JEF distributions in \jetptrange{60}{80} for jet pairs with radii [0.05, 0.1], [0.1, 0.15], [0.15, 0.2], and [0.2, 0.25], in Pb--Pb collisions at $\sqrt{s_{\rm NN}} = 5.02~\mathrm{TeV}$. The lines represent the statistical uncertainties, and the boxes represent the systematic uncertainties.}
\label{fig:jef_PbPb_data}
\end{figure}

The fully corrected JEF distributions for central Pb--Pb collisions (0--10\% centrality class)  in the interval \jetptrange{60}{80} for jet pairs with radii of [0.05, 0.1], [0.1, 0.15], [0.15, 0.2], and [0.2, 0.25] are shown in Fig.~\ref{fig:jef_PbPb_data}. The measurement in Pb--Pb collisions is restricted to higher jet \pt and smaller radius pairs in order to keep the background fluctuations under control and to ensure a stable unfolding. The systematic uncertainties are larger for wider jets, but remain sufficiently small to allow a meaningful comparison among different jet radii. A similar trend to that observed in pp collisions is seen here. The distributions are peaked at $\dpt=0$ and fall steeply toward larger \dpt. The steepness is found to increase smoothly with increasing radius, with a flatter behavior for the smallest jet-radius pair, $R_{\mathrm {jet_1}}={0.05}$ and $R_{\mathrm {jet_2}}={0.10}$, and the steepest fall for the largest pair, $R_{\mathrm {jet_1}}={0.2}$ and $R_{\mathrm {jet_2}}={0.25}$. The distributions for the two largest-radius jet pairs are observed to be closer to each other than other sets of radius values, and they exhibit similar shapes.
\subsection{Modification of JEF in Pb--Pb collisions}
\begin{figure}[!htb]
\centering
   \includegraphics[scale=0.35]{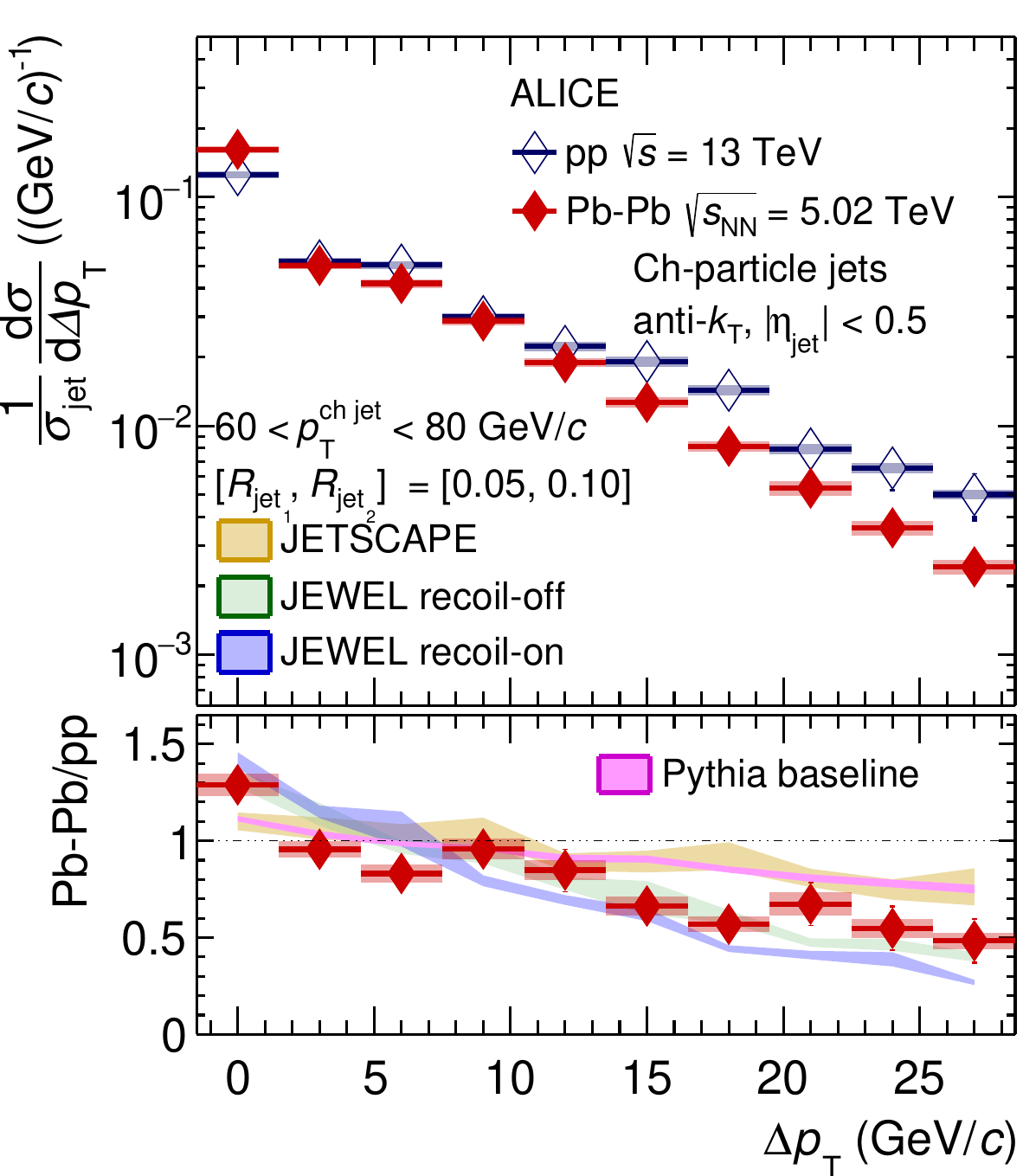}
    \includegraphics[scale=0.35]{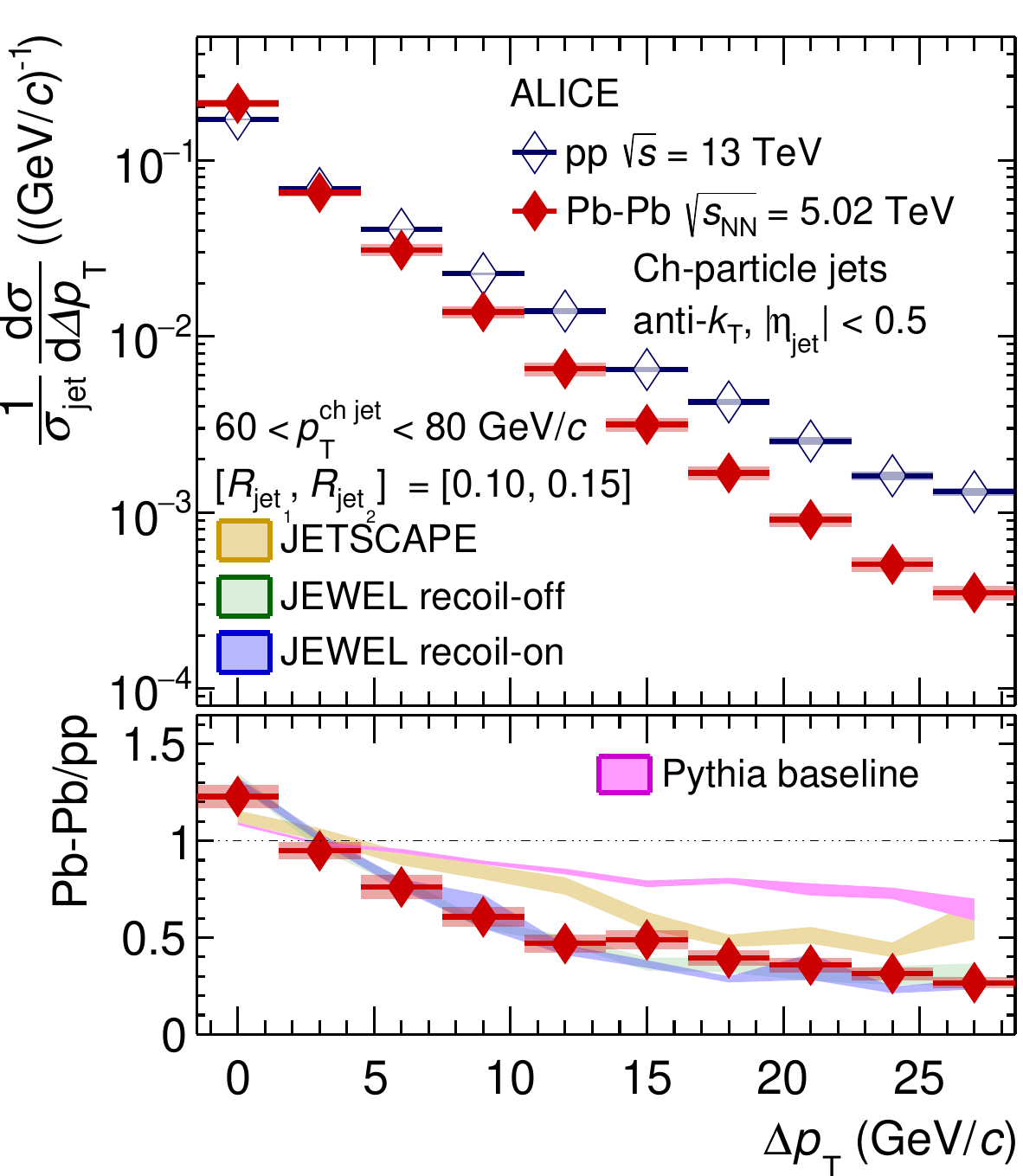}
       \includegraphics[scale=0.35]{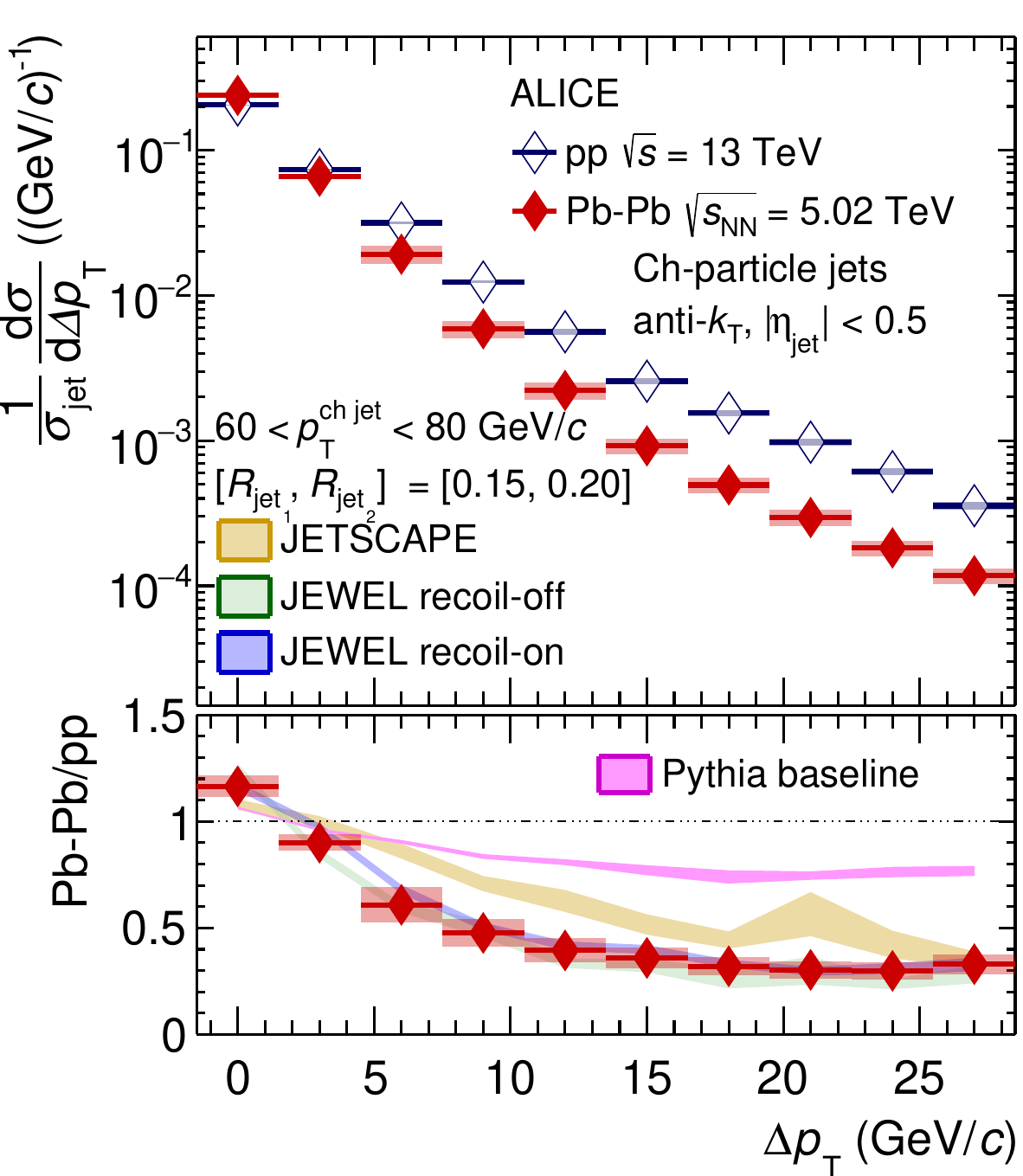}
    \includegraphics[scale=0.35]{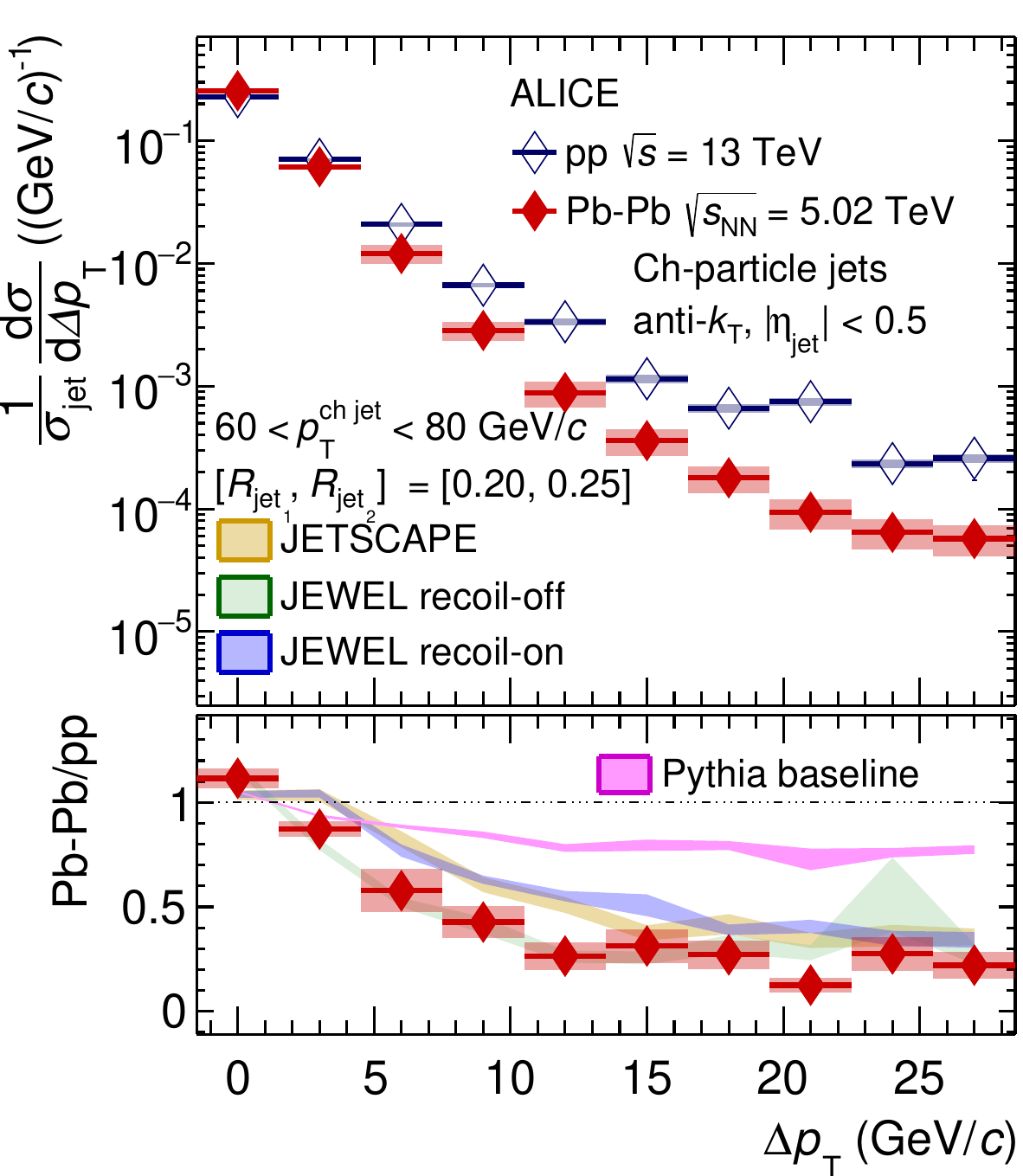}
    \caption{JEF distributions in the \jetptrange{60}{80} interval for pp collisions at $\sqrt{s}=13~\mathrm{TeV}$ and Pb--Pb collisions at $\sqrt{s_{\rm NN}}=5.02~\mathrm{TeV}$ are shown for jet pairs with radii [0.05, 0.1] (top left), [0.1, 0.15] (top right), [0.15, 0.2] (bottom left), and [0.2, 0.25] (bottom right). The bottom panels show the ratio of JEF in Pb--Pb collisions to pp collisions, compared to JETSCAPE and JEWEL (with recoil on and off) models. The Pythia baseline is shown as the magenta band in the bottom panels.}
\label{fig:jef_modificaiton_PbPb_data}
\end{figure}

Figure~\ref{fig:jef_modificaiton_PbPb_data} shows a comparison of the JEF measurement in pp collisions at $\sqrt{s}=13~$TeV and Pb--Pb collisions at $\sqrt{s_{\rm NN}}=5.02~$TeV for different jet pairs. The top panel compares the JEF distributions, illustrating that the Pb--Pb distributions are steeper than those in pp collisions, with the effect being more pronounced for the largest radius pair. This trend is more clearly seen in the bottom panel, which shows the ratio of the JEF measurement in Pb--Pb to that in pp collisions, together with comparisons to different models. The relative uncertainties are assumed to be uncorrelated between collision systems. A clear relative suppression is observed for $\dpt \geq 5~\GeVc$ for all jet radii. The largest radius pair exhibits a steeper suppression than the smallest radius pair, indicating a relative ``narrowing'' of jets.

To further understand jet–medium interactions, the observed relative suppression is compared with predictions from several theoretical models. The magenta band in the bottom panels, labeled as the ``Pythia baseline'', shows the ratio of JEF in pp collisions at $\sqrt{s}=5.02~$TeV to pp collisions at $\sqrt{s}=13~$TeV, illustrating the expected dependence on the collision energy. The significance of the data points at the largest \dpt relative to the Pythia baseline is evaluated, considering both statistical and systematic uncertainties. The last four data points are included in this assessment. The suppression at each of these $\Delta p_{\rm T}$ values is approximately $2~\sigma$ for the smallest-radius jet pairs [0.05, 0.1], increasing to about $5~\sigma$ for the largest-radius pairs [0.2, 0.25].

These models incorporate different implementations of the microscopic properties of the medium, its evolution, and the jet–medium interaction. In JETSCAPE~\cite{Putschke:2019yrg}, a medium-modified parton shower is described by MATTER and LBT. The bulk medium initial conditions are setup; using the TRENTO model~\cite{Moreland:2014oya} followed by (2+1)-dimensional free-streaming pre-equilibrium evolution~\cite{Liu:2015nwa} and the QGP medium evolution is described by the VISHNU (2+1)-dimensional hydrodynamics~\cite{Song:2007ux}.
In the JEWEL event generator~\cite{Zapp:2012ak,Zapp:2013vla}, the hard-scattering event is generated by PYTHIA 6, followed by a medium-modified shower which takes into account elastic scattering and splittings, including formation time. The medium is modeled as a boost-invariant, longitudinally expanding ideal quark–gluon gas and hadronization is handled by PYTHIA 6~\cite{Zapp:2012ak}. The parameter set from Ref.~\cite{KunnawalkamElayavalli:2017hxo}, which is appropriate for the kinematic range of this measurement, is used. The medium partons recoiling after interactions with the jet constituents can either be discarded from the event (recoils off) or allowed to hadronize together with the jet (recoils on). For the JEWEL prediction with recoil on, the contribution from thermal partons is subtracted using the momenta of the medium partons stored in the JEWEL output and a subtraction algorithm that is based on the constituent-subtraction algorithm~\cite{Milhano:2022kzx}.

The models show overall agreement with the measured Pb--Pb/pp ratios. For the JETSCAPE model, the agreement is better in pp collisions (as shown in Fig.~\ref{fig:jef_pp_60_80}) than in Pb--Pb collisions, while the Pb--Pb/pp ratio is still reasonably well reproduced. In contrast, JEWEL in pp shows overall good agreement, with some deviations in the tails of the distributions for jet pairs with small radii. For Pb--Pb/pp ratios, JEWEL with recoil on and off performs similarly for jet pairs with small radii. However, for jet pairs with larger radii ($R_{\rm jet} \geq 0.20$), larger deviations are observed for the recoil-on case, while the recoil-off case shows better agreement. This behavior is also observed in other observables, such as the $R_{\rm AA}$ double ratio~\cite{ALICE:2023waz}.

\begin{figure}[!htb]
\centering
   \includegraphics[scale=0.5]{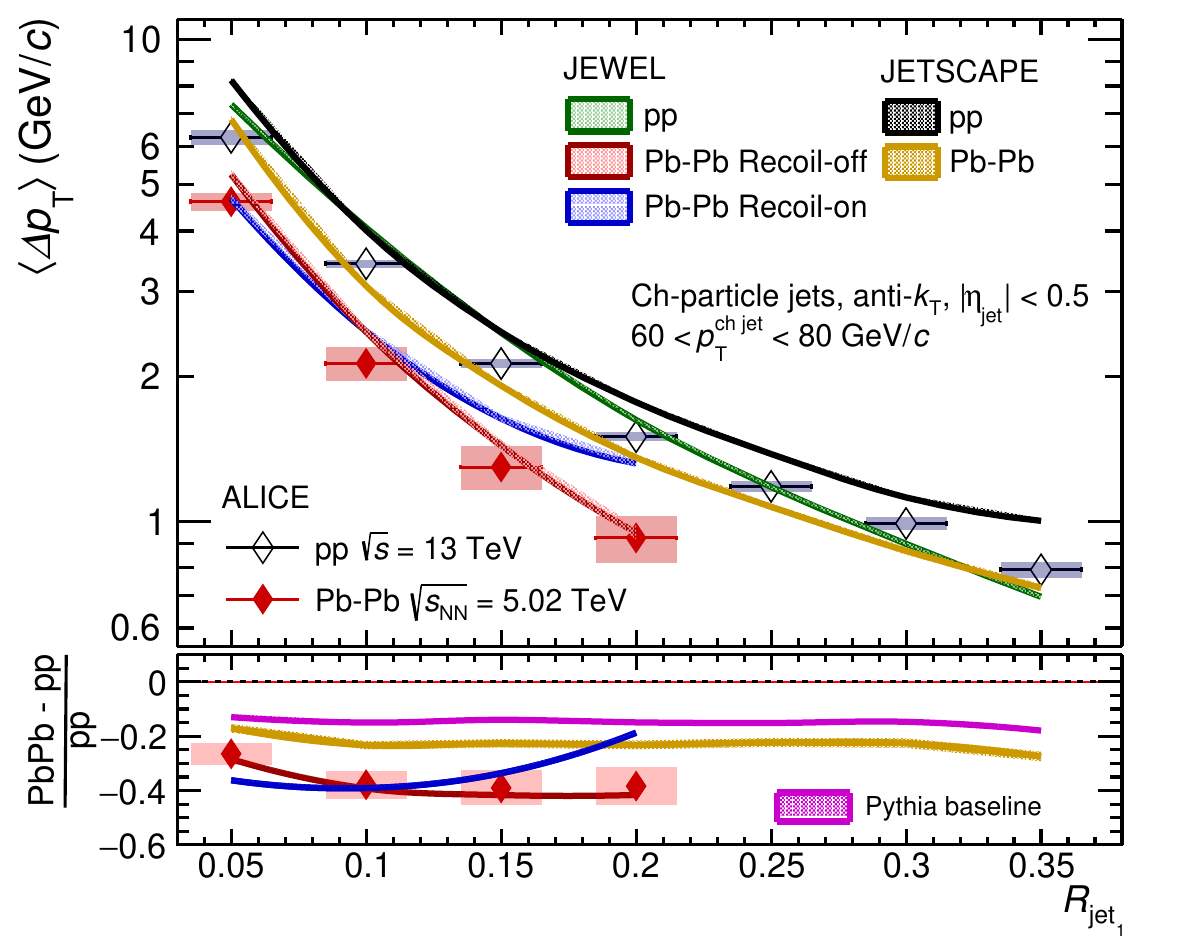}
    \caption{ Top: Comparison of the mean energy flow for jet pairs with radii between \Rjet{0.05} and \Rjet{0.4} (\Rjet{0.25} for Pb--Pb) in the {\mbox{\jetptrange{60}{80}}} interval in ALICE pp (Pb--Pb) collisions, compared to models. In the model calculations $\sqrt{s}=13~\mathrm{TeV}$ is used for pp collisions and $\sqrt{s_{\mathrm{NN}}}=5.02~\mathrm{TeV}$ for Pb--Pb collisions. Bottom: Ratio of the relative difference of Pb--Pb with respect to pp, compared to JETSCAPE and JEWEL (with recoil on and off). The Pythia baseline is shown as the magenta band in the bottom panel. }
\label{fig:jef_modificaiton_PbPb_data_jewel}
\end{figure}

\subsubsection*{Modification of mean jet-energy flow \dptmean}
To further investigate the evolution of the jet-energy flow with jet radius in Pb--Pb collisions and its modification with respect to pp, the mean energy flow in Pb--Pb is reported and compared to that in pp and to different models in Fig.~\ref{fig:jef_modificaiton_PbPb_data_jewel}. The mean energy flow exhibits a similar trend to that observed in pp collisions, indicating that the jet-energy flow remains concentrated in the core. However, differences in its evolution with jet radius are observed, as clearly illustrated in the bottom panel, which shows the relative difference of Pb--Pb with respect to pp. The ratio decreases with increasing $R$ in the second bin and then becomes flat, indicating a narrowing of the jet energy profile in the reported kinematic range.

JEWEL in both pp and Pb–Pb collisions, for both recoil-on and recoil-off configurations, provides a better description of the mean \dpt results compared to JETSCAPE. The magenta band in the bottom panel, labeled as the ``Pythia baseline'', shows the relative difference of pp at $\sqrt{s}=5.02~$TeV with respect to pp at $\sqrt{s}=13~$TeV, to illustrate the expected dependence on the collision energy. For the last three data points, the suppression relative to the PYTHIA baseline lies in the range $3.5$–$4.5~\sigma$. While JETSCAPE predicts a suppression, the constant trend is disfavoured by the data. The ratio is best reproduced by JEWEL with recoils switched off (red curve in the bottom panels). In this case, the energy and momentum transferred from hard partons to recoil partons do not appear in the event record. This is a reasonable approximation for hard jet structure observables, but it may not be accurate for jet observables that include low-momentum particles, particularly when using larger jet radii. JEWEL with recoil shows a clear increase in the relative jet-energy flow between Pb--Pb and pp, which is not seen in the data.

The radial jet profile modification measured by the CMS Collaboration~\cite{CMS:2013lhm} in central Pb--Pb collisions also shows a suppression at smaller angles ($\leq 0.2$), followed by an enhancement of the momentum flow at larger angles. Follow-up studies of the jet momentum profile by CMS~\cite{CMS:2018zze} demonstrate that the energy redistribution at large angles is primarily carried by low-momentum particles. The redistribution of energy at large angles observed in the jet-shape measurements cannot be confirmed in the present analysis due to the limited $R_{\rm jet, low}$ reach in Pb--Pb collisions. Therefore, measurements with extended jet radii would be valuable to assess this trend and enable more stringent comparisons with model predictions.

\section{Conclusions}
\label{sec:conclusions}
The JEF is measured in pp collisions at $\sqrt{s}=13$~TeV and in central Pb--Pb collisions (0--10\% centrality class) at $\sqrt{s_{\rm NN}}=5.02$~TeV with ALICE. In pp collisions, the JEF exhibits a clear ordering with jet radius, with the energy concentrated in the jet core and decreasing smoothly toward larger radii, consistent with collimated vacuum parton showers and well described by several MC generators, with some deviations observed in the tails of the distributions. This establishes a robust baseline for the Pb--Pb measurement.

In Pb--Pb collisions, the JEF distributions are significantly steeper than in pp, and a suppression at large $\Delta p_{\rm T}$ is observed, which increases with jet radius. This behavior indicates a narrowing of the jet energy profile due to medium-induced modifications. Comparisons to theoretical models show that while the overall trends are qualitatively reproduced, the description depends sensitively on the treatment of the medium response. Extending future measurements to larger jet radii will enable the investigation of a possible redistribution of radiation at large angles. Furthermore, extending the measurement to lower and higher jet momenta will allow the study of the \pt dependence of the jet-energy flow. 

%%%%% acknowledgements
\newenvironment{acknowledgement}{\relax}{\relax}
\begin{acknowledgement}
\section*{Acknowledgements}
% Version: 2026-06-03

The ALICE Collaboration would like to thank all its engineers and technicians for their invaluable contributions to the construction of the experiment and the CERN accelerator teams for the outstanding performance of the LHC complex.
The ALICE Collaboration gratefully acknowledges the resources and support provided by all Grid centres and the Worldwide LHC Computing Grid (WLCG) collaboration.
The ALICE Collaboration acknowledges the following funding agencies for their support in building and running the ALICE detector:
A. I. Alikhanyan National Science Laboratory (Yerevan Physics Institute) Foundation (ANSL), State Committee of Science and World Federation of Scientists (WFS), Armenia;
Austrian Academy of Sciences, Austrian Science Fund (FWF): [M 2467-N36] and Nationalstiftung f\"{u}r Forschung, Technologie und Entwicklung, Austria;
Ministry of Communications and High Technologies, National Nuclear Research Center, Azerbaijan;
Rede Nacional de Física de Altas Energias (Renafae), Financiadora de Estudos e Projetos (Finep), Funda\c{c}\~{a}o de Amparo \`{a} Pesquisa do Estado de S\~{a}o Paulo (FAPESP) and The Sao Paulo Research Foundation  (FAPESP), Brazil;
Bulgarian Ministry of Education and Science, within the National Roadmap for Research Infrastructures 2020-2027 (object CERN), Bulgaria;
Ministry of Education of China (MOEC) , Ministry of Science \& Technology of China (MSTC) and National Natural Science Foundation of China (NSFC), China;
Ministry of Science and Education and Croatian Science Foundation, Croatia;
Centro de Aplicaciones Tecnol\'{o}gicas y Desarrollo Nuclear (CEADEN), Cubaenerg\'{\i}a, Cuba;
Ministry of Education, Youth and Sports of the Czech Republic, Czech Republic;
The Danish Council for Independent Research | Natural Sciences, the VILLUM FONDEN and Danish National Research Foundation (DNRF), Denmark;
Helsinki Institute of Physics (HIP), Finland;
Commissariat \`{a} l'Energie Atomique (CEA) and Institut National de Physique Nucl\'{e}aire et de Physique des Particules (IN2P3) and Centre National de la Recherche Scientifique (CNRS), France;
Bundesministerium f\"{u}r Forschung, Technologie und Raumfahrt (BMFTR) and GSI Helmholtzzentrum f\"{u}r Schwerionenforschung GmbH, Germany;
National Research, Development and Innovation Office, Hungary;
Department of Atomic Energy Government of India (DAE), Department of Science and Technology, Government of India (DST), University Grants Commission, Government of India (UGC) and Council of Scientific and Industrial Research (CSIR), India;
National Research and Innovation Agency - BRIN, Indonesia;
Istituto Nazionale di Fisica Nucleare (INFN), Italy;
Japanese Ministry of Education, Culture, Sports, Science and Technology (MEXT) and Japan Society for the Promotion of Science (JSPS) KAKENHI, Japan;
Consejo Nacional de Ciencia (CONACYT) y Tecnolog\'{i}a, through Fondo de Cooperaci\'{o}n Internacional en Ciencia y Tecnolog\'{i}a (FONCICYT) and Direcci\'{o}n General de Asuntos del Personal Academico (DGAPA), Mexico;
Nederlandse Organisatie voor Wetenschappelijk Onderzoek (NWO), Netherlands;
The Research Council of Norway, Norway;
Pontificia Universidad Cat\'{o}lica del Per\'{u}, Peru;
Ministry of Science and Higher Education, National Science Centre and WUT ID-UB, Poland;
Korea Institute of Science and Technology Information and National Research Foundation of Korea (NRF), Republic of Korea;
Ministry of Education and Scientific Research, Institute of Atomic Physics, Ministry of Research and Innovation and Institute of Atomic Physics and Universitatea Nationala de Stiinta si Tehnologie Politehnica Bucuresti, Romania;
Ministerstvo skolstva, vyskumu, vyvoja a mladeze SR, Slovakia;
National Research Foundation of South Africa, South Africa;
Swedish Research Council (VR) and Knut \& Alice Wallenberg Foundation (KAW), Sweden;
European Organization for Nuclear Research, Switzerland;
Suranaree University of Technology (SUT), National Science and Technology Development Agency (NSTDA) and National Science, Research and Innovation Fund (NSRF via PMU-B B05F650021), Thailand;
Turkish Energy, Nuclear and Mineral Research Agency (TENMAK), Turkey;
National Academy of  Sciences of Ukraine, Ukraine;
Science and Technology Facilities Council (STFC), United Kingdom;
National Science Foundation of the United States of America (NSF) and United States Department of Energy, Office of Nuclear Physics (DOE NP), United States of America.
In addition, individual groups or members have received support from:
Czech Science Foundation (grant no. 23-07499S), Czech Republic;
FORTE project, reg.\ no.\ CZ.02.01.01/00/22\_008/0004632, Czech Republic, co-funded by the European Union, Czech Republic;
European Research Council (grant no. 101220549), European Union;
Deutsche Forschungs Gemeinschaft (DFG, German Research Foundation) ``Neutrinos and Dark Matter in Astro- and Particle Physics'' (grant no. SFB 1258), Germany;
CONVECS project, CUP C97H23001700002 FESR 2021-2027 program, Italy.

\end{acknowledgement}

%%%%%%%% Bibliography (In case of using bibtex generate the bbl requested by arXiv)
%\bibliographystyle{elsarticle-num}
\bibliographystyle{utphys}   % Remember we use title in the biblio
\bibliography{main}
%\input {bibliography.tex}  

%%%%%%%%% appendix with author list
\newpage
\appendix
\section{The ALICE Collaboration}
\label{app:collab}
% ALICE Collaboration author list for 2026-06-03
\begin{flushleft} 
\small

D.A.H.~Abdallah\,\orcidlink{0000-0003-4768-2718}\,$^{\rm 134}$, 
I.J.~Abualrob\,\orcidlink{0009-0005-3519-5631}\,$^{\rm 112}$, 
S.~Acharya\,\orcidlink{0000-0002-9213-5329}\,$^{\rm 49}$, 
K.~Agarwal\,\orcidlink{0000-0001-5781-3393}\,$^{\rm II,}$$^{\rm 23}$, 
G.~Aglieri Rinella\,\orcidlink{0000-0002-9611-3696}\,$^{\rm 32}$, 
L.~Aglietta\,\orcidlink{0009-0003-0763-6802}\,$^{\rm 24}$, 
N.~Agrawal\,\orcidlink{0000-0003-0348-9836}\,$^{\rm 25}$, 
Z.~Ahammed\,\orcidlink{0000-0001-5241-7412}\,$^{\rm 132}$, 
S.~Ahmad\,\orcidlink{0000-0003-0497-5705}\,$^{\rm 15}$, 
Z.~Akbar$^{\rm 79}$
% ...

\section*{Affiliation Notes}

$^{\rm I}$ Deceased\\
$^{\rm II}$ Also at: INFN Trieste, Trieste, Italy\\
$^{\rm III}$ Also at: Fondazione Bruno Kessler (FBK), Trento, Italy\\
$^{\rm IV}$ Also at: Czech Technical University in Prague, Prague, Czech Republic\\
$^{\rm V}$ Also at: Instituto de Fisica da Universidade de Sao Paulo\\
$^{\rm VI}$ Also at: Dipartimento DET del Politecnico di Torino, Turin, Italy\\
$^{\rm VII}$ Also at: University College of Dublin, Dublin, Ireland\\
$^{\rm VIII}$ Also at: Department of Applied Physics, Aligarh Muslim University, Aligarh, India\\
$^{\rm IX}$ Also at: Institute of Theoretical Physics, University of Wroclaw, Wroclaw, Poland\\
$^{\rm X}$ Also at: Facultad de Ciencias, Universidad Nacional Aut\'{o}noma de M\'{e}xico, Mexico City, Mexico\\

\section*{Collaboration Institutes}

$^{1}$ A.I. Alikhanyan National Science Laboratory (Yerevan Physics Institute) Foundation, Yerevan, Armenia\\
$^{2}$ AGH University of Krakow, Cracow, Poland\\
$^{3}$ Bogolyubov Institute for Theoretical Physics, National Academy of Sciences of Ukraine, Kyiv, Ukraine\\
$^{4}$ Bose Institute, Department of Physics  and Centre for Astroparticle Physics and Space Science (CAPSS), Kolkata, India\\
$^{5}$ California Polytechnic State University, San Luis Obispo, California, United States\\
$^{6}$ Central China Normal University, Wuhan, China\\
$^{7}$ Centro de Aplicaciones Tecnol\'{o}gicas y Desarrollo Nuclear (CEADEN), Havana, Cuba\\
$^{8}$ Centro de Investigaci\'{o}n y de Estudios Avanzados (CINVESTAV), Mexico City and M\'{e}rida, Mexico\\
$^{9}$ Chicago State University, Chicago, Illinois, United States\\
$^{10}$ China Nuclear Data Center, China Institute of Atomic Energy, Beijing, China\\
$^{11}$ China University of Geosciences, Wuhan, China\\
$^{12}$ Chungbuk National University, Cheongju, Republic of Korea\\
$^{13}$ Comenius University Bratislava, Faculty of Mathematics, Physics and Informatics, Bratislava, Slovak Republic\\
$^{14}$ Creighton University, Omaha, Nebraska, United States\\
$^{15}$ Department of Physics, Aligarh Muslim University, Aligarh, India\\
$^{16}$ Department of Physics, Pusan National University, Pusan, Republic of Korea\\
$^{17}$ Department of Physics, Sejong University, Seoul, Republic of Korea\\
$^{18}$ Department of Physics, University of California, Berkeley, California, United States\\
$^{19}$ Department of Physics, University of Oslo, Oslo, Norway\\
$^{20}$ Department of Physics and Technology, University of Bergen, Bergen, Norway\\
$^{21}$ Dipartimento di Fisica, Universit\`{a} di Pavia, Pavia, Italy\\
$^{22}$ Dipartimento di Fisica dell'Universit\`{a} and Sezione INFN, Cagliari, Italy\\
$^{23}$ Dipartimento di Fisica dell'Universit\`{a} and Sezione INFN, Trieste, Italy\\
$^{24}$ Dipartimento di Fisica dell'Universit\`{a} and Sezione INFN, Turin, Italy\\
$^{25}$ Dipartimento di Fisica e Astronomia dell'Universit\`{a} and Sezione INFN, Bologna, Italy\\
$^{26}$ Dipartimento di Fisica e Astronomia dell'Universit\`{a} and Sezione INFN, Catania, Italy\\
$^{27}$ Dipartimento di Fisica e Astronomia dell'Universit\`{a} and Sezione INFN, Padova, Italy\\
$^{28}$ Dipartimento di Fisica `E.R.~Caianiello' dell'Universit\`{a} and Gruppo Collegato INFN, Salerno, Italy\\
$^{29}$ Dipartimento DISAT del Politecnico and Sezione INFN, Turin, Italy\\
$^{30}$ Dipartimento di Scienze MIFT, Universit\`{a} di Messina, Messina, Italy\\
$^{31}$ Dipartimento Interateneo di Fisica `M.~Merlin' and Sezione INFN, Bari, Italy\\
$^{32}$ European Organization for Nuclear Research (CERN), Geneva, Switzerland\\
$^{33}$ Faculty of Electrical Engineering, Mechanical Engineering and Naval Architecture, University of Split, Split, Croatia\\
$^{34}$ Faculty of Nuclear Sciences and Physical Engineering, Czech Technical University in Prague, Prague, Czech Republic\\
$^{35}$ Faculty of Physics, Sofia University, Sofia, Bulgaria\\
$^{36}$ Faculty of Science, P.J.~\v{S}af\'{a}rik University, Ko\v{s}ice, Slovak Republic\\
$^{37}$ Faculty of Technology, Environmental and Social Sciences, Bergen, Norway\\
$^{38}$ Frankfurt Institute for Advanced Studies, Johann Wolfgang Goethe-Universit\"{a}t Frankfurt, Frankfurt, Germany\\
$^{39}$ Fudan University, Shanghai, China\\
$^{40}$ Gauhati University, Department of Physics, Guwahati, India\\
$^{41}$ Helmholtz-Institut f\"{u}r Strahlen- und Kernphysik, Rheinische Friedrich-Wilhelms-Universit\"{a}t Bonn, Bonn, Germany\\
$^{42}$ Helsinki Institute of Physics (HIP), Helsinki, Finland\\
$^{43}$ High Energy Physics Group,  Universidad Aut\'{o}noma de Puebla, Puebla, Mexico\\
$^{44}$ Horia Hulubei National Institute of Physics and Nuclear Engineering, Bucharest, Romania\\
$^{45}$ HUN-REN Wigner Research Centre for Physics, Budapest, Hungary\\
$^{46}$ Indian Institute of Technology Bombay (IIT), Mumbai, India\\
$^{47}$ Indian Institute of Technology Indore, Indore, India\\
$^{48}$ INFN, Laboratori Nazionali di Frascati, Frascati, Italy\\
$^{49}$ INFN, Sezione di Bari, Bari, Italy\\
$^{50}$ INFN, Sezione di Bologna, Bologna, Italy\\
$^{51}$ INFN, Sezione di Cagliari, Cagliari, Italy\\
$^{52}$ INFN, Sezione di Catania, Catania, Italy\\
$^{53}$ INFN, Sezione di Padova, Padova, Italy\\
$^{54}$ INFN, Sezione di Pavia, Pavia, Italy\\
$^{55}$ INFN, Sezione di Torino, Turin, Italy\\
$^{56}$ INFN, Sezione di Trieste, Trieste, Italy\\
$^{57}$ Inha University, Incheon, Republic of Korea\\
$^{58}$ Institute for Gravitational and Subatomic Physics (GRASP), Utrecht University/Nikhef, Utrecht, Netherlands\\
$^{59}$ Institute of Experimental Physics, Slovak Academy of Sciences, Ko\v{s}ice, Slovak Republic\\
$^{60}$ Institute of Physics, Homi Bhabha National Institute, Bhubaneswar, India\\
$^{61}$ Institute of Physics of the Czech Academy of Sciences, Prague, Czech Republic\\
$^{62}$ Institute of Space Science (ISS), Bucharest, Romania\\
$^{63}$ Institut f\"{u}r Kernphysik, Johann Wolfgang Goethe-Universit\"{a}t Frankfurt, Frankfurt, Germany\\
$^{64}$ Instituto de Ciencias Nucleares, Universidad Nacional Aut\'{o}noma de M\'{e}xico, Mexico City, Mexico\\
$^{65}$ Instituto de F\'{i}sica, Universidade Federal do Rio Grande do Sul (UFRGS), Porto Alegre, Brazil\\
$^{66}$ Instituto de F\'{\i}sica, Universidad Nacional Aut\'{o}noma de M\'{e}xico, Mexico City, Mexico\\
$^{67}$ iThemba LABS, National Research Foundation, Somerset West, South Africa\\
$^{68}$ Jeonbuk National University, Jeonju, Republic of Korea\\
$^{69}$ Korea Institute of Science and Technology Information, Daejeon, Republic of Korea\\
$^{70}$ Laboratoire de Physique Subatomique et de Cosmologie, Universit\'{e} Grenoble-Alpes, CNRS-IN2P3, Grenoble, France\\
$^{71}$ Lawrence Berkeley National Laboratory, Berkeley, California, United States\\
$^{72}$ Lund University Department of Physics, Division of Particle Physics, Lund, Sweden\\
$^{73}$ Marietta Blau Institute, Vienna, Austria\\
$^{74}$ Nagasaki Institute of Applied Science, Nagasaki, Japan\\
$^{75}$ Nara Women{'}s University (NWU), Nara, Japan\\
$^{76}$ National Centre for Nuclear Research, Warsaw, Poland\\
$^{77}$ National Institute of Science Education and Research, Homi Bhabha National Institute, Jatni, India\\
$^{78}$ National Nuclear Research Center, Baku, Azerbaijan\\
$^{79}$ National Research and Innovation Agency - BRIN, Jakarta, Indonesia\\
$^{80}$ Niels Bohr Institute, University of Copenhagen, Copenhagen, Denmark\\
$^{81}$ Nikhef, National institute for subatomic physics, Amsterdam, Netherlands\\
$^{82}$ Nuclear Physics Group, STFC Daresbury Laboratory, Daresbury, United Kingdom\\
$^{83}$ Nuclear Physics Institute of the Czech Academy of Sciences, Husinec-\v{R}e\v{z}, Czech Republic\\
$^{84}$ Oak Ridge National Laboratory, Oak Ridge, Tennessee, United States\\
$^{85}$ Ohio State University, Columbus, Ohio, United States\\
$^{86}$ Physics department, Faculty of science, University of Zagreb, Zagreb, Croatia\\
$^{87}$ Physics Department, Panjab University, Chandigarh, India\\
$^{88}$ Physics Department, University of Jammu, Jammu, India\\
$^{89}$ Physics Program and International Institute for Sustainability with Knotted Chiral Meta Matter (WPI-SKCM$^{2}$), Hiroshima University, Hiroshima, Japan\\
$^{90}$ Physikalisches Institut, Eberhard-Karls-Universit\"{a}t T\"{u}bingen, T\"{u}bingen, Germany\\
$^{91}$ Physikalisches Institut, Ruprecht-Karls-Universit\"{a}t Heidelberg, Heidelberg, Germany\\
$^{92}$ Physik Department, Technische Universit\"{a}t M\"{u}nchen, Munich, Germany\\
$^{93}$ Politecnico di Bari and Sezione INFN, Bari, Italy\\
$^{94}$ Research Division and ExtreMe Matter Institute EMMI, GSI Helmholtzzentrum f\"ur Schwerionenforschung GmbH, Darmstadt, Germany\\
$^{95}$ Saga University, Saga, Japan\\
$^{96}$ Saha Institute of Nuclear Physics, Homi Bhabha National Institute, Kolkata, India\\
$^{97}$ School of Physics and Astronomy, University of Birmingham, Birmingham, United Kingdom\\
$^{98}$ Secci\'{o}n F\'{\i}sica, Departamento de Ciencias, Pontificia Universidad Cat\'{o}lica del Per\'{u}, Lima, Peru\\
$^{99}$ SUBATECH, IMT Atlantique, Nantes Universit\'{e}, CNRS-IN2P3, Nantes, France\\
$^{100}$ Sungkyunkwan University, Suwon City, Republic of Korea\\
$^{101}$ Suranaree University of Technology, Nakhon Ratchasima, Thailand\\
$^{102}$ Technical University of Ko\v{s}ice, Ko\v{s}ice, Slovak Republic\\
$^{103}$ The Henryk Niewodniczanski Institute of Nuclear Physics, Polish Academy of Sciences, Cracow, Poland\\
$^{104}$ The University of Texas at Austin, Austin, Texas, United States\\
$^{105}$ Universidad Aut\'{o}noma de Sinaloa, Culiac\'{a}n, Mexico\\
$^{106}$ Universidade de S\~{a}o Paulo (USP), S\~{a}o Paulo, Brazil\\
$^{107}$ Universidade Estadual de Campinas (UNICAMP), Campinas, Brazil\\
$^{108}$ Universidade Federal do ABC, Santo Andre, Brazil\\
$^{109}$ Universitatea Nationala de Stiinta si Tehnologie Politehnica Bucuresti, Bucharest, Romania\\
$^{110}$ University of Cape Town, Cape Town, South Africa\\
$^{111}$ University of Derby, Derby, United Kingdom\\
$^{112}$ University of Houston, Houston, Texas, United States\\
$^{113}$ University of Jyv\"{a}skyl\"{a}, Jyv\"{a}skyl\"{a}, Finland\\
$^{114}$ University of Kansas, Lawrence, Kansas, United States\\
$^{115}$ University of Liverpool, Liverpool, United Kingdom\\
$^{116}$ University of Science and Technology of China, Hefei, China\\
$^{117}$ University of Silesia in Katowice, Katowice, Poland\\
$^{118}$ University of South-Eastern Norway, Kongsberg, Norway\\
$^{119}$ University of Tennessee, Knoxville, Tennessee, United States\\
$^{120}$ University of the Witwatersrand, Johannesburg, South Africa\\
$^{121}$ University of Tokyo, Tokyo, Japan\\
$^{122}$ University of Tsukuba, Tsukuba, Japan\\
$^{123}$ Universit\"{a}t M\"{u}nster, Institut f\"{u}r Kernphysik, M\"{u}nster, Germany\\
$^{124}$ Universit\'{e} Clermont Auvergne, CNRS/IN2P3, LPC, Clermont-Ferrand, France\\
$^{125}$ Universit\'{e} de Lyon, CNRS/IN2P3, Institut de Physique des 2 Infinis de Lyon, Lyon, France\\
$^{126}$ Universit\'{e} de Strasbourg, CNRS, IPHC UMR 7178, F-67000 Strasbourg, France, Strasbourg, France\\
$^{127}$ Universit\'{e} Paris-Saclay, Centre d'Etudes de Saclay (CEA), IRFU, D\'{e}partment de Physique Nucl\'{e}aire (DPhN), Saclay, France\\
$^{128}$ Universit\'{e}  Paris-Saclay, CNRS/IN2P3, IJCLab, Orsay, France\\
$^{129}$ Universit\`{a} degli Studi di Foggia, Foggia, Italy\\
$^{130}$ Universit\`{a} del Piemonte Orientale, Vercelli, Italy\\
$^{131}$ Universit\`{a} di Brescia, Brescia, Italy\\
$^{132}$ Variable Energy Cyclotron Centre, Homi Bhabha National Institute, Kolkata, India\\
$^{133}$ Warsaw University of Technology, Warsaw, Poland\\
$^{134}$ Wayne State University, Detroit, Michigan, United States\\
$^{135}$ Yale University, New Haven, Connecticut, United States\\
$^{136}$ Yildiz Technical University, Istanbul, Turkey\\
$^{137}$ Yonsei University, Seoul, Republic of Korea\\
$^{138}$ Affiliated with an institute formerly covered by a cooperation agreement with CERN\\
$^{139}$ Affiliated with an international laboratory covered by a cooperation agreement with CERN.\\

\end{flushleft} 
  %%%%%%% done by webmaster team
\end{document}